\documentclass[12pt]{article}
\usepackage{amsthm,amsmath,amsfonts,amssymb}
\usepackage{graphicx,psfrag,epsf}
\usepackage{bmpsize}
\usepackage{enumerate}
\usepackage{natbib}
\usepackage{algorithm}
\usepackage{algorithmic}
\usepackage[title,toc,titletoc]{appendix}
\usepackage{url} % not crucial - just used below for the URL 
\usepackage{hyperref}
\newcommand{\blind}{0}

\usepackage[dvipsnames]{xcolor}
\usepackage{bm}
\usepackage{bbm}

\newcommand{\betab}{\bm{\beta}}

\newcommand{\thetab}{\bm{\theta}}  %% theta vector
\newcommand{\db}{\bm{d}}

\newcommand{\xb}{\bm{x}}    %% bold x vector
\newcommand{\Xb}{\bm{X}}
\newcommand{\zb}{\bm{z}}
\newcommand{\lambdab}{\bm{\lambda}}

\newcommand{\Deltab}{\bm{\Delta}}

\newcommand*\diff{\mathop{}\!\textrm{d}}

\DeclareMathOperator*{\argmin}{argmin}

\begin{document}

\def\spacingset#1{\renewcommand{\baselinestretch}%
{#1}\small\normalsize} \spacingset{1}

%%%%%%%%%%%%%%%%%%%%%%%%%%%%%%%%%%%%%%%%%%%%%%%%%%%%%%%%%%%%%%%%%%%%%%%%%%%%%%

\if0\blind
{
  \title{\bf Black Box Variational Bayesian Model Averaging}
  \author{Vojtech Kejzlar\\
    Department of Mathematics and Statistics, Skidmore College\\
    Shrijita Bhattacharya\\
    Department of Statistics and Probability, Michigan State University\\
    Mookyong Son\\
    Department of Statistics and Probability, Michigan State University\\
    and \\
    Tapabrata Maiti \\
    Department of Statistics and Probability, Michigan State University}
  \maketitle
} \fi

\if1\blind
{
  \bigskip
  \bigskip
  \bigskip
  \begin{center}
    {\LARGE\bf Black Box Variational Bayesian Model Averaging}
\end{center}
  \medskip
} \fi

\bigskip
\begin{abstract} For many decades now, Bayesian Model Averaging (BMA) has been a popular framework to systematically account for model uncertainty  that arises in situations when multiple competing models are available to describe the same or similar physical process. The implementation of this framework, however, comes with a multitude of practical challenges including posterior approximation via Markov Chain Monte Carlo and numerical integration. We present a Variational Bayesian Inference approach to BMA as a viable alternative to the standard solutions which avoids many of the aforementioned pitfalls. The proposed method is ``black box'' in the sense that it can be readily applied to many models with little to no model-specific derivation. We illustrate the utility of our variational approach on a suite of examples and discuss all the necessary implementation details. Fully documented \texttt{Python} code with all the examples is provided as well.
\end{abstract}

\noindent%
\vfill

\spacingset{1.45} % DON'T change the spacing!
\section{Introduction}

The existence of several competing models to solve the same or similar problem is a common scenario across scientific applications. One typically encounters a slew of candidate models during standard regression analysis with multiple predictors. Another widely familiar example is a numerical weather prediction with multitudes of forecasting models available. The routine practice in this situation is to select a single model and then make inference based on this model which ignores a major component of uncertainty - model uncertainty \citep{Leamer1978}. Bayesian model averaging (BMA) is the natural Bayesian framework to systematically account for uncertainty due to several competing models.

For any quantity of interest $\Deltab$, such as a future observation or an effect size, the BMA posterior density $p(\Deltab|\db)$ corresponds to the mixture of posterior densities of the individual models $p(\Deltab|\db,M)$ weighted by their posterior model probabilities $p(M|\db)$:
\begin{equation} \label{eqn:posteriorBMA}
p(\Deltab|\db) 
= \sum_{M \in \mathcal{M}} p(\Deltab|\db,M) p(M|\db),
\end{equation}
where $\mathcal{M}$ denotes the space of all models, $\db = (d_1, \cdots,d_n)$ are given datapoints, and $d_i = (x_i,y_i)$ for $i = 1,\dots , n$ are input-observation pairs. Note, if the space of models is $\mathcal{M}=\{M_1, \cdots,M_K\}$, then the formula in \eqref{eqn:posteriorBMA} can be equivalenty written as $p(\Deltab|\db) 
= \sum_{k=1}^K p(\Deltab|\db,M_k) p(M_k|\db)$ which is a more commonly used notation for BMA. We stick to the notation in \eqref{eqn:posteriorBMA} to facilitate the developments in section \ref{chapter:BBVI}

The posterior probability of a model $M$ is given by a simple application of the Bayes' theorem:
\begin{equation} \label{eqn:posteriorsMmodel:BMA}
p(M|\db)
= \frac{p(\db|M)p(M)}{\sum_{M' \in \mathcal{M}}p(\db|M') p(M')}.
\end{equation}

Due to the mixture form of the density \eqref{eqn:posteriorBMA}, determining these probabilities is the key to successful implementation of the BMA framework. To do so, one first needs to assign a suitable prior probability $p(M)$ that $M$ is the \textit{true model} (assuming there is one such, among the models considered). \cite{BMA} notes that,
\begin{quote}
    \textit{When there is little prior information about the relative plausibility of the models considered, the assumption that all models are equally likely a priory is a reasonable ``neutral" choice.}
\end{quote}
One can, nevertheless, choose informative prior distributions when prior information about the likelihood of each model is available. Eliciting an informative prior is a non-trivial task, but \cite{Madigan1995} provide some guidance in the context of graphical models that can be applied in other settings as well.

The second component of model's posterior probability is the model's marginal likelihood, also known as model evidence,
\begin{equation} \label{eqn:evidencel:BMA}
p(\db|M) =\int p(\db|\thetab_{M}, M)p(\thetab_{M}|M)\diff \thetab_{M},
\end{equation}
where $\thetab_{M}$ is the set of model-specific parameters ($\thetab_M = (\betab, \sigma)$ in regression problems), $p(\thetab_{M}|M)$ is their prior distribution, and $p(\db|\thetab_{M}, M)$ is the model's data likelihood. The evaluation of model evidence is one of the main reasons why BMA becomes computationally challenging in practice, because a closed form solution is available only in special scenarios for the exponential family of distributions with conjugate priors, and thus the integral \eqref{eqn:evidencel:BMA} requires approximation. Some problem-specific algorithms have been developed for direct sampling from BMA posterior density \eqref{eqn:posteriorBMA} such as the Markov chain Monte Carlo (MCMC) model composition for linear regression models (MC$^3$) \citep{Raftery1997}.

A vast body of literature was produced over the past 30 years on the topic of model evidence approximation, with the simplest approach being the Monte Carlo (MC) integration. The advantage of MC integration is in the method's ease of implementation, however, one typically needs to generate a large number of samples from prior distribution to achieve reasonable convergence. A popular improvement to the simple MC integration is the harmonic mean estimator which makes the use of samples from posterior distribution of model parameters and therefore converges more quickly. On the other hand, it can be unstable, and it tends to overestimate the evidence (see \cite{Raftery2007, Lenk2009}). A large class of statistically efficient estimators is based on importance sampling that relies on draws from an importance density which approximates the joint posterior density of model parameters. However, a poor choice of importance density may lead to a huge loss of efficiency. See \cite{Neal2001}, \cite{Freil2008}, and \cite{Pajor2017} for some examples of estimators with importance sampling. Another classical method is the Laplace approximation. This corresponds to a second order Taylor expansion of the log-likelihood around its maximum, which makes the likelihood normal. Laplace method is efficient for well behaved likelihoods. We refer the reader to \cite{KassRaftery1995, ARDIA2012} and \cite{Friel2012} for a complete survey of popular approximation methods for the model evidence. Additionally, the more recently proposed Nested Sampling algorithm by \cite{NestedSampling} and expanded by \cite{MultiNest} provides another alternative to the aforementioned approaches.

Here we want to point out that the definition of BMA relies on the assumption that the true model which represents the physical reality is within the models being considered (i.e. $\mathcal{M}$-closed setting). BMA can lead to misleading results when the true model is not included (i.e. $\mathcal{M}$-open setting). For instance, a scenario with two models -- one mediocre and one "perfect" almost everywhere with a large deviation from the truth at a single point of the input space -- will typically result in the selection of the mediocre model. Similarly, BMA can also lead to a suboptimal performance under model misspecification \citep{Clarke2003, Masegosa2020}. Using BMA in the $\mathcal{M}$-open setting additionally creates a logical tension between interpreting $p(M|\db)$ and $p(M)$ as probabilities of $M$ being the true model and knowing that the true model is not in $\mathcal{M}$. One can perhaps reconcile this tension by considering $p(M|\db)$ and $p(M)$ as the probabilities of $M$ being a useful description of physical reality. In what follows, we will assume that the reader is comfortable with assigning a prior over $\mathcal{M}$, even in the $\mathcal{M}$-open setting. We refer to \cite{Bernardo1994} for a detailed discussion about the conceptual differences between the $\mathcal{M}$-closed and $\mathcal{M}$-open settings and to \cite{Fragoso} for a recent survey of BMA methodology. A decision-theoretic approach to account for model uncertainty in $\mathcal{M}$-open setting is presented in \cite{Iversen2013}. Recently, \cite{BAND} proposed a model-mixing approach for the case when the list of models considered does not contain the true model. Both of these methods address the inadequacy of BMA in $\mathcal{M}$-open setting by not considering models as an extension of the parameter space.

Despite its conceptual and computational challenges, BMA has a long history of use in both natural sciences and humanities because of a superior predictive performance that is theoretically guaranteed \citep{Bernardo1994}. \cite{Geweke} introduced BMA in economics and later in other fields such as political and social sciences. See the recent review on the use of BMA in Economics by \cite{Steel2020}. BMA has also been applied to the medical sciences \citep{med-Vi14, med-SCH16}, 
ecology and evolution \citep{ecol-Si14,ecol-Ho15}, 
genetics \citep{gen-Vis11, gen-Wen15}, machine learning \citep{ML-Mer11, ML-Hernandez2018, Dunson2020}, and lately in nuclear physics \citep{Neufcourt2019, Neufcourt2020, Neufcourt2020a, Kejzlar2020}.

In this paper, we present a Variational Bayesian Inference (VBI) approach to BMA. VBI is a useful alternative to the sampling-based approximation via MCMC that approximates a target density through optimization. Statisticians and computer scientists (starting with \cite{Peterson, Jordan1999}) have been widely using variational techniques because they tend to be faster and easier to scale to massive datasets. Our method is based on the variational inference algorithm with reparametrization gradients developed by \cite{Titsias14} and \cite{Kucukelbir2017} which can be applied to many models with minimum additional derivations. The proposed approach, which we shall call the black box variational BMA (VBMA), is a one step procedure that simultaneously approximates model evidences and posterior distributions of individual models while enjoying all the advantages (and disadvantages) of VBI. Here we note that this is not the first time a VBI is used in the context of BMA. For instance, \cite{latoucheVBMA} developed a variational Bayesian approach specifically for averaging of graphon functions, and \cite{VBMA2014} use VBI and BMA for audio source separation. However, the VBMA is a general algorithm that can be applied directly to a wide class of models including Bayesian neural networks, generalized linear models, and Gaussian process models.

\subsection{Outline of this paper}
In section 2, we provide a brief overview of VBI and derive our proposed VBMA algorithm. Then, in section 3, we present a collection of examples that include standard linear regression, logistic regression, and Bayesian model selection. To fully showcase the computational benefits of VBMA, we consider Gaussian process models for residuals of separation energies of atomic nuclei. We compare VBMA with direct sampling BMA via MC$^3$ and with MCMC posterior approximation and evidence computed using MC integration. A fully documented Python code with our algorithm and examples is available at \url{https://github.com/kejzlarv/BBVBMA}. Lastly, in section 4, we discuss the pros and cons of VBMA and provide a list of sensible machine learning applications for the proposed methodology.

\section{BMA via Variational Bayesian Inference} \label{chapter:VBI}
\subsection{Variational Bayesian Inference}
VBI strives to approximate a target posterior distribution through optimization. One first considers a family of distributions $q(\thetab|\lambdab)$, indexed by a variational parameter $\lambdab$, over the space of model parameters and subsequently finds a member of this family $q^*$ closest to the posterior distribution $p(\thetab|\db)$. The simplest variational family is the mean-field family which assumes independence of all the components in $\thetab$ but many other families of variational distributions exist; see \cite{MAL001,pmlr-v38-hoffman15, Ranganath:2016, Tran2015, Tran:2017, pmlr-v37-rezende15,  NIPS2016_ddeebdee, Kucukelbir2017,  fortunato,   NIPS2017_6c1da886, JMLR:v22:19-1028, normalizing, pmlr-v108-weilbach20a}. The recent work of \cite{pmlr-v130-ambrogioni21a} and the references therein provide a detailed discussion of these class of variational families and their associated implementation challenges. The approximate distribution $q^*$ is chosen to minimize the  Kullback-Leibler (KL) divergence of $q(\thetab|\lambdab)$ from $p(\thetab|\db)$:
\begin{equation}\label{eqn:VI_def}
q^* = \argmin_{q(\thetab|\lambdab)} KL(q(\thetab|\lambdab)||p(\thetab|\db)).
\end{equation}

Finding $q^*$ is done in practice by maximizing an equivalent objective function \citep{Jordan1999}, the \textit{evidence lower bound (ELBO)}:
\begin{equation}\label{eqn:ELBO}
	\mathcal{L}(q) = \mathbb{E}_{ q(\thetab|\lambdab)}\bigg[\log p(\db, \thetab) - \log q(\thetab|\lambdab)\bigg].
\end{equation}

The ELBO is the sum between the negative KL divergence of the variational distribution from the true posterior distribution
and the log of the marginal data distribution $p(\db)$. The term $\log p(\db)$ is constant with respect to $q(\thetab|\lambdab)$. It is also a lower bound on $\log p(\db)$ for any choice of $q(\thetab|\lambdab)$. ELBO can be optimized via standard coordinate- or gradient-ascent methods. However, these techniques are inefficient for large datasets, and so it has become common practice to use the stochastic gradient ascent (SGA) algorithm. SGA updates $\lambdab$ at the $t^{th}$ iteration according to
\begin{equation}\label{eqn:StochasticUpdate}
    \lambdab_{t+1} \leftarrow \lambdab_{t} + \rho_t \tilde{l}(\lambdab_t),
\end{equation}
where $\tilde{l}(\lambdab)$ is a realization of the random variable $\tilde{\mathcal{L}}(\lambdab)$ which is an unbiased estimate of the gradient $\nabla_{\lambdab} \mathcal{L}(\lambdab)$. 

Let us now assume that $\log p(\db, \thetab)$ and $\log q(\thetab|\lambdab)$ are differentiable functions with respect to $\thetab$, and that the random variable $\thetab$ can be reparametrized using a differentiable transformation $t(\zb, \lambdab)$ of an auxiliary variable $\zb$ so that $\zb \sim \psi(\zb)$ and $\thetab = t(\zb, \lambdab)$ imply $\thetab \sim q(\thetab|\lambdab)$. It is assumed that $\psi(\zb)$ exists in a standard form so that any parameter mean vector is set to zero and scale parameters are set to one. For example, if we consider a real valued $\theta$ with normal variation family $q(\theta|\mu, \sigma^2)$, then $t(z, (\mu, \sigma)) = z \sigma + \mu$ and $z\sim \text{Normal}(0,1)$. Note that the variational parameters are part of the transformation and not the auxiliary distribution. The gradient of the ELBO can be then expressed as the following expectation with respect to the auxiliary distribution $\psi(\zb)$ \citep{Titsias14, Kucukelbir2017}:
\begin{equation}\label{eqn:VBI:ELBOdelta}
\nabla_{\lambdab} \mathcal{L}(q) = \mathbb{E}_{ \psi(\zb)}\bigg[\nabla_{\thetab}(\log p(\db, \thetab) - \log q(\thetab|\lambdab)) \times \nabla_{\lambdab}t(\zb, \lambdab)\bigg].
\end{equation}

The expectation \eqref{eqn:VBI:ELBOdelta} does not have a closed form in general, nevertheless, one can use $S$ samples from $\psi(\zb)$ to construct its unbiased MC estimate for the SGA \eqref{eqn:StochasticUpdate}
\begin{align}\label{eqn:VBI:ELBOdelta_estimate}
\tilde{l}(\lambdab) = \frac{1}{S}\sum^{S}_{s=1}\bigg[\nabla_{\thetab}(\log p(\db, t(\zb[s], \lambdab)) - \log q(t(\zb[s], \lambdab)|\lambdab)) \times \nabla_{\lambdab}t(\zb[s], \lambdab)\bigg],
\end{align}
where $\zb[s] \sim \psi(\zb)$. Since that the differentiability assumptions and the reparametrization trick allows the use of autodifferentiation to take gradients, the method is \textit{black box} in nature \citep{Kucukelbir2017}. The disadvantage of reparametrization gradient is that it requires differentiable models, i.e, models with no discrete variables. One can use the so called score gradient \citep{Ranganath14} for models with discrete variables which is also black box in nature, however, the variance of the gradient estimates can be large and lead to unreliable results \citep{OBBVI}. The estimate \eqref{eqn:VBI:ELBOdelta_estimate} can be conveniently used in the SGA algorithm which converges to a local maximum of $\mathcal{L}(\lambdab)$ (global for $\mathcal{L}(\lambdab)$ concave \citep{bottou-97}) when the learning rate $\rho_t$ follows the Robbins-Monro conditions \citep{robbins1951}
\begin{equation}\label{eqn:VBI:RMconditions}
\sum_{t = 1}^{\infty}  \rho_t = \infty, \hspace{1cm} \sum_{t = 1}^{\infty}  \rho^2_t < \infty.
\end{equation}

Choosing an optimal learning rate $\rho_t$ can be challenging in practice. Ideally, one would want the rate to be small in situations where MC estimates of the ELBO gradient are erratic (large variance) and large when the MC estimates are relatively stable (small variance). The elements of variational parameter $\lambda$ can also differ in scale, and the selected learning rate should accommodate these varying, potentially small, scales. The ever increasing abundance of stochastic optimization in machine learning applications spawned development of numerous algorithms for element-wise adaptive scale learning rates. We use the Adam algorithm \citep{Adam} which is a popular and easy-to-implement adaptive rate algorithm. However, there are many other frequently used algorithms such as the AdaGrad \citep{AdaGrad}, the ADADELTA \citep{Zeiler2012}, or the RMSprop \citep{Tieleman2012}. The step size associated with Adam is kept constant throughout the paper. However, as pointed out in \cite{pmlr-v80-shazeer18a}, one may achieve a better performance by making use of linear ramp-up followed by some form of decay \citep{NIPS2017_3f5ee243}. In the Supplement, we provide the results based on RMSprop for comparison. We did not observe significant differences between RMSprop and Adam for the examples in section \ref{chapter:Applications}.

Below, we extend  the standard VBI that approximates a distribution of model parameters to a scenario where a distribution over the model space needs to be also approximated.

\subsection{Black Box Variational BMA}
\label{chapter:BBVI}
For any quantity of interest $\Deltab$, such as a future observation or an effect size, the BMA posterior density $p(\Deltab|\db)$ corresponds to the mixture of posterior densities of the individual models $p(\Deltab|\db,M)$ weighted by their posterior model probabilities $p(M|\db)$ as in  equations \eqref{eqn:posteriorBMA}, \eqref{eqn:posteriorsMmodel:BMA}, and \eqref{eqn:evidencel:BMA}.

In order to facilitate variational inference, we reformulate the problem of BMA as follows
\begin{equation}
\label{e:posterior}
 p(\Deltab|\db)=\int p(\Deltab|\db, M, \thetab_{M})p(M,\thetab_{M}|\db) \diff \mu(M, \thetab_{M})
\end{equation}
where $\mu$ is the product measure of counting and Lebesgue. Note, the expression \eqref{e:posterior} indeed summarizes the equations \eqref{eqn:posteriorBMA}, \eqref{eqn:posteriorsMmodel:BMA} and \eqref{eqn:evidencel:BMA} in one step. In practice, the most difficult quantity to compute is $p(M,\thetab_{M}|\db)$. We shall now consider the joint posterior distribution of the model $M$  and its corresponding parameter $\thetab_{M}$ as our parameter of interest. Note, the above notation allows for the dependence of $\thetab$ on the model $M$. This is needed as the indexing parameter of each model could differ in dimension, distribution, etc. As explained in Section \ref{chapter:VBI}, there has been a plethora of literature in using variational inference to obtain the posterior distribution $\thetab$ for a given model.  In this section, we adapt the variational inference to obtain the joint distribution of the model and the parameter together in one stroke.

We next assume a variational approximation to the posterior distribution $p(M,\thetab_{M}|\db)$ of the form $q(M,\thetab_{M}|\lambdab_{M})=q(M)q(\thetab_{M}|M, \lambdab_{M})$, where $q(M)$ is the variational model weight of model $M$ and $q(\thetab_M|\lambdab_M)$ is the variational distribution of $\thetab_M$ under model $M$ and is indexed by its corresponding variational parameter $\lambdab_M$.  Note that we treat $(M,\thetab_M)$ as a random variable which takes on the values $(m,\thetab_m)$ for varying values of $m \in \mathcal{M}=\{M_1, \cdots, M_K\}$. The density of this random variable is given by $p(m,\thetab_m|\db)$ under the true posterior, and by $q(m)q(\thetab_m|\lambdab_m)$ under the variational posterior. Here $q(m)$ for $m \in \mathcal{M}$ can be any categorical distribution satisfying $\sum_{m\in \mathcal{M}}q(m)=1$. For each $m \in \mathcal{M}$, $q(\thetab_m|\lambdab_m)$ can be any parametric distribution indexed by the parameters $\lambdab_m$. Possible choices of $q(\thetab_m|\lambdab_m)$ include but are not restricted to mean field variational family of the form $q(\thetab_m|\lambdab_m)=\prod_{i} q(\theta_{m}^i|\lambda_m^i)$. We additionally assume that $\thetab_{M}$ can be reparametrized using a differentiable transformation $t(\zb_{M}, \lambdab_{M})$ of an auxiliary variable $\zb_{M}$ so that $\zb_{M} \sim \psi(\zb_{M})$ and $\thetab_{M} = t(\zb_{M}, \lambdab_{M})$. We avoid the inherent dependence of $t(\cdot)$ on $M$ to simplify the notation.

Thus, the optimal variational distribution $q^*$ is given by 
\begin{equation}\label{eqn:KL}
q^* = \argmin_{q} KL(q(M,\thetab_{M}|\lambdab_M)||p(M,\thetab_{M}|\db)).
\end{equation}
Again, in the KL expression above, we assume that $M$ is a random variable whose values are the individual models in the model space $\mathcal{M}$. As explained in section \ref{chapter:VBI}, finding $q^*$ is obtained in practice by maximizing an equivalent objective function \citep{Jordan1999}, the ELBO
\begin{equation}
	\mathcal{L}(q)= \mathbb{E}_{q(M,\thetab_{M}| \lambdab_M)}\bigg[\log p(\db, M,\thetab_{M}) - \log q(M,\thetab_{M}|\lambdab_{M})\bigg],
\end{equation}
this time, subject to constraint $\sum_{M \in \mathcal{M}} q(M) = 1$. Since $q(\thetab_M|\lambdab_M)$ is a parametric family indexed by the parameters $\lambdab_M$, it is indeed a valid density function. However, since the categorical distribution $q(M)$ has freely varying parameters, the constraint $\sum_{M\in \mathcal{M}}q(M)=1$ is imposed. To accommodate the constraint, using Lagrange multipliers, we optimize
$$\mathcal{L}(q) = \mathbb{E}_{q(M,\thetab_{M}|\lambdab_M)}[\log p(\db, M,\thetab_{M})-\log q(M,\thetab_{M}|\lambdab_{M})] -\varrho(\sum_{M \in \mathcal{M}} q(M)-1).$$
The ELBO can be simplified further
\begin{align*}
&\mathbb{E}_{q(M,\thetab_{M}|\lambda_{M})}[\log p(\db, M,\thetab_{M})-\log q(M,\thetab_{M}|\lambdab_{M})] -\varrho(\sum_{M \in \mathcal{M}} q(M)-1)\\
&=\mathbb{E}_{q(M,\thetab_{M}|\lambda_{M})}[\log p(\db|M,\thetab_{M})+\log p(\thetab_{M}|M)+\log p(M)\\
&\enskip \qquad \qquad \qquad-\log q(\thetab_{M}|M,\lambdab_{M})-\log q(M)]-\varrho(\sum_{M \in \mathcal{M}} q(M)-1)\\
&=\sum_{M \in \mathcal{M}} q(M)\mathbb{E}_{q(\thetab_{M}|M, \lambdab_M)}[\log p(\db|M,\thetab_{M})+\log p(\thetab_{M}|M)+\log p(M) \\
&\enskip \enskip \qquad \qquad \qquad \qquad \qquad -\log q(\thetab_{M}|M, \lambdab_M)-\log q(M)] - 
\varrho(\sum_{M \in \mathcal{M}} q(M)-1).
\end{align*}

\noindent Since the parameters $q(M)$ for $M \in \mathcal{M}$ do not depend on the variational parameters $\lambdab_M$, therefore $\nabla_{\lambda_M}\varrho(\sum_{M \in \mathcal{M}} q(M)-1)=0$. Thus, one obtains $\nabla_{\lambda_M} \mathcal{L}(q) =q(M)\mathcal{G}_M$ where
\begin{align*}
{\mathcal{G}_M}&=\mathbb{E}_{\psi(\zb_M)}[\nabla_{\thetab_M} (\log p(\db|M,\thetab_{M})+\log p(\thetab_M|M)-\log q(\thetab_M|M, \lambdab_M)) \times \nabla_{\lambdab_M} t(\zb_M, \lambdab_M)].
\end{align*}

To estimate the quantity $\mathcal{G}_M$, we can generate multiple samples from the distribution $\psi(\zb_M)$ and then use the MC estimate
\begin{align*}
  \widehat{\mathcal{G}}_M=\frac{1}{S}\sum_{s=1}^S [\nabla_{\thetab_M} &(\log p(\db|M, t(\zb_M[s], \lambdab_M))+\log p(t(\zb_M[s], \lambdab_M)|M) \\
  &-\log q(t(\zb_M[s], \lambdab_M)|M, \lambdab_M)) \times \nabla_{\lambdab_M} t(\zb_M[s], \lambdab_M)] .
\end{align*}

To derive the update of $q(M)$, note that
\begin{align*}
\nabla_{q(M)} \mathcal{L}(q) &= \underbrace{\mathbb{E}_{q(\thetab_M|M, \lambdab_M)}[(\log p(\db|M,\thetab_{M})+\log p(\thetab_M|M)-\log q(\thetab_M|M, \lambdab_M))]}_{\mathcal{L}_M}\\
&+\log p(M)-\log q(M)-1 - \varrho,
\end{align*}
where $\mathcal{L}_{M}$ is nothing but the ELBO under a fixed model $M$.
Equating the above derivative to 0, we get a closed form expression for $q(M)$ as 
$$q(M)=\exp(\mathcal{L}_M+\log p(M)-1 - \varrho) \propto \exp(\mathcal{L}_M+\log p(M)).$$

It only remains to generate the quantity $\mathcal{L}_{M}$, which we get again by multiple samples from the distribution $\psi(\zb_M)$ and then use the MC estimate
$$\widehat{\mathcal{L}}_M=\frac{1}{S}\sum_{s=1}^S [\log p(\db|M,t(\zb_M[s], \lambdab_M))+\log p(t(\zb_M[s], \lambdab_M)|M)-\log q(t(\zb_M[s], \lambdab_M)|M, \lambdab_{M})].$$
This allows us to get Algorithm~\ref{alg:BBVBMA} for VBMA.

    \begin{algorithm}[!t]
    \caption{Black Box Variational BMA \label{alg:BBVBMA}}
    \begin{algorithmic}
    \STATE Start with an initial choice of $(\lambdab_M, q(M))_{M \in \mathcal{M}}$ and a learning rate $\rho$.
    \REPEAT
    \STATE By generating $\zb_M[1], \cdots, \zb_M[S]$ from $\psi(\zb_M)$, calculate
    \begin{align*}
        \widehat{\mathcal{G}}_M=\frac{1}{S}\sum_{s=1}^S [\nabla_{\thetab_M} &(\log p(\db|M, t(\zb_M[s], \lambdab_M))+\log p(t(\zb_M[s], \lambdab_M)|M) \\
  &-\log q(t(\zb_M[s], \lambdab_M)|M, \lambdab_M)) \times \nabla_{\lambdab_M} t(\zb_M[s], \lambdab_M)] 
    \end{align*}
\STATE Update $\lambdab_M$ as 
$$\lambdab_M=\lambdab_M + \rho q(M)\widehat{\mathcal{G}}_M$$
\STATE Using the already generated $\zb_M[1], \cdots, \zb_M[S]$, calculate
\begin{align*}
\widehat{\mathcal{L}}_M=\frac{1}{S}\sum_{s=1}^S &[\log p(\db|M,t(\zb_M[s], \lambdab_M))+\log p(t(\zb_M[s], \lambdab_M)|M) \\
&-\log q(t(\zb_M[s], \lambdab_M)|M, \lambdab_{M})]
\end{align*}
\STATE Update $q(M)$ as
$$\widetilde{q}(M) = \exp(\widehat{\mathcal{L}}_M+\log p(M))$$
and $q(M) = \widetilde{q}(M) / \sum_{M \in \mathcal{M}} \widetilde{q}(M)$.
\UNTIL{Convergence of $\widehat{\mathcal{L}}(q)$ where}
$$\widehat{\mathcal{L}}(q)=\sum_{M \in \mathcal{M}}q(M) \widehat{\mathcal{L}}_{M}$$
\end{algorithmic}
    \end{algorithm}
  
\subsubsection{Implementation Details and Variational Families}\label{chapter:Parametrization}

The general form of VBMA algorithm allows the user to select the variational family that is most appropriate for the problem at hand. As we noted in Section \ref{chapter:VBI}, there is a vast pool of candidate families that vary by their expressiveness and ability to capture complex structure of unknown parameters many of which can be used in reparametrization gradients. In the subsequent applications, we shall consider mean-field variational families with normal distributions for real valued variables and log-normal distributions for positive variables. Despite its simplicity, the mean-field variational family can approximate a wide class of posteriors and is good enough to achieve consistency for the variational posterior for a wide class of models \citep{VIth, ZhangGao, BHATTACHARYA2021151}. Moreover, all the strictly positive variational parameters $\lambda$ will be transformed as
\begin{equation}\label{eqn:reparametrization}
    \Tilde{\lambda} = \log (e^{\lambda} - 1)
\end{equation}
to avoid constrained optimization. See Appendix~\ref{sec:appendix:reparametrization} for the details on the reparametrization of normal and log-normal mean-field families.

Besides the choice of suitable variational family for VBMA, another practical consideration needs to be made regarding the updates of variational parameter given by $\lambdab_M=\lambdab_M + \rho q(M)\widehat{\mathcal{G}}_M$. Since each step directly depends on the variational approximation of the posterior model probabilities $q(M)$, the updates can be computationally unstable unless the ELBO of each individual model is close to convergence. We therefore recommend setting $q(M) := 1/K$, where $K$ is the number of models considered, until the variation approximation of the posterior model probabilities stabilizes. Additionally, we recommend to compute the final values of $\widetilde{q}(M)$, and $q(M)$ respectively, as an average of the last several hundred iterations of the Algorithm \eqref{alg:BBVBMA} for a greater reliability of the estimates.

\section{Examples}\label{chapter:Applications}
Below, we provide a suite of illustrative real data examples to demonstrate how VBMA serves as a viable alternative to approximate the BMA posterior distribution. First, we analyze the U.S. crime data  under the standard linear regression model. Second, we consider a logistic regression model for a heart disease dataset. We also show that VBMA provides a convenient solution to Bayesian model selection with Bayes factors. To fully showcase the computational benefits of VBMA, we study Gaussian process models for the residuals of separation energies of atomic nuclei where the standard MCMC-based implementation is challenging in practice.  Each of the examples looks at a situation with several competing models without any prior knowledge of which is better; thus we set the prior model weights to be uniform over the model space. All the reparametrization gradients in the following applications were obtained using the autodifferentiation engine in \texttt{Python} package \texttt{PyTorch} \citep{Torch}.

\subsection{Bayesian Linear Regression}
In this example, we compare VBMA with the MCMC algorithm MC$^3$ using the aggregated crime data on 47 U.S. states of \cite{vandaele1978participation} which has been considered by \cite{Raftery1997} to illustrate the efficiency of BMA in regression scenario with a multitude of candidate models. For simplicity, we concentrate only on a minimal subset of 3 out of 15 predictors of the crime rate and following \cite{Raftery1997}, we log transformed all the continuous variables (predictors were also centered). 

Given the response variable $y$, we consider models of the form 
\begin{equation}\label{eqn_lin_reg}
    y = \beta_0 + \sum_{j=1}^{p}\beta_j x_j  + \epsilon,
\end{equation}
where $x_1,\dots, x_p$ is a subset of a set of candidate predictors $x_1, \dots, x_k$. In this specific example, we consider three predictors: $x_1$ corresponding to the percentage of males age 14-24, $x_2$ corresponds to the probability of imprisonment, and $x_3$ contains the mean years of schooling in the state. We assign $\epsilon$ a normal distribution with mean zero and precision $\phi$. The $\epsilon$'s are assumed to be independent for distinct cases. For the parameters in each model \eqref{eqn_lin_reg}, we use Zellner's g-prior \citep{Zellner1986, Raftery1997}
\begin{align*}
    \phi &\propto 1/ \phi, \\
    \beta_0 &\propto 1, \\
    \beta_1, \dots \beta_p & \propto N(0, g (\Xb^{'}\Xb)^{-1}/ \phi),
\end{align*}
where $g=n$ and $\Xb$ is the design matrix. Zellner's g-prior is one of the most popular conjugate Normal-Gamma prior distributions for linear models that is convenient and provides Bayesian computation with marginal likelihoods that can be evaluated analytically.

\subsubsection{Results}

\begin{table}[h!]
\caption{\label{tab:reg_prob} The top four linear regression models of the crime data according to their posterior model probabilities. The star indicates the inclusion of predictor in the model and the model ID is provided for easier referencing. Comparison between the VBMA and the MC based averaging is shown.}
\begin{center}
\begin{tabular}{|c|c|c|c|c|r|r|}
\cline{2-7}
\multicolumn{1}{l|}{} & \multicolumn{4}{c|}{\textbf{Inclusion}} & \multicolumn{2}{c|}{$p(M|\db)$} \\ \hline
\multicolumn{1}{|l|}{\textbf{Model}} & \multicolumn{1}{c|}{\textbf{Intercept}} & \multicolumn{1}{c|}{$\bm{x_1}$} & \multicolumn{1}{c|}{$\bm{x_2}$} & \multicolumn{1}{c|}{$\bm{x_3}$} & \multicolumn{1}{c|}{\textbf{MC$^3$}} & \multicolumn{1}{c|}{\textbf{VBMA}} \\ \hline
0 & * &  & * &  &  $0.58$  & $0.57$  \\
1 & * &  & * & * & 0.17 & 0.15 \\
2 & * & * &  * &  &  $ 0.11$ & $0.11$ \\
3 & * & * & * & * & 0.07 & 0.05 \\
\hline
\end{tabular}
\end{center}
\end{table}

Table \ref{tab:reg_prob} shows the estimates of model posterior probabilities obtained with VBMA and through the MCMC algorithm MC$^3$ for the top four models. The VBMA results are based on a pre-training sequence of $500$ iterations with the model probabilities set to 1/8 and $200$ iterations of updating according to Algorithm \ref{alg:BBVBMA}. Ten MC samples from the variational distributions were used to estimate the ELBO gradient. The displayed probabilities were determined as the average over the last $100$ iterations of the algorithm to ensure stability of the estimates. The MC$^3$ results were computed with \texttt{R} package \texttt{BAS} \citep{ML-Mer11}. Clearly, the VBMA based values closely match the MC$^3$ with small deviations for the models with lower posterior probabilities. However, this difference does not dramatically impact the data analysis.

\begin{figure}[h!]
\caption{Posterior distributions for predictor slopes based on VBMA (first row) and MC$^3$ (second row). Namely, $\beta_1$ corresponds to the percentage of males 14-24, $\beta_2$ to the probability of imprisonment, and $\beta_3$ to the mean years of schooling. The density is scaled so that the maximum of the density is equal to $\mathbb{P}(\beta \ne 0|\db)$. The spike corresponds to to $\mathbb{P}(\beta = 0|\db)$.\label{fig:post_betas_lin}}
	\begin{center}		
	\includegraphics[width=0.9\textwidth]{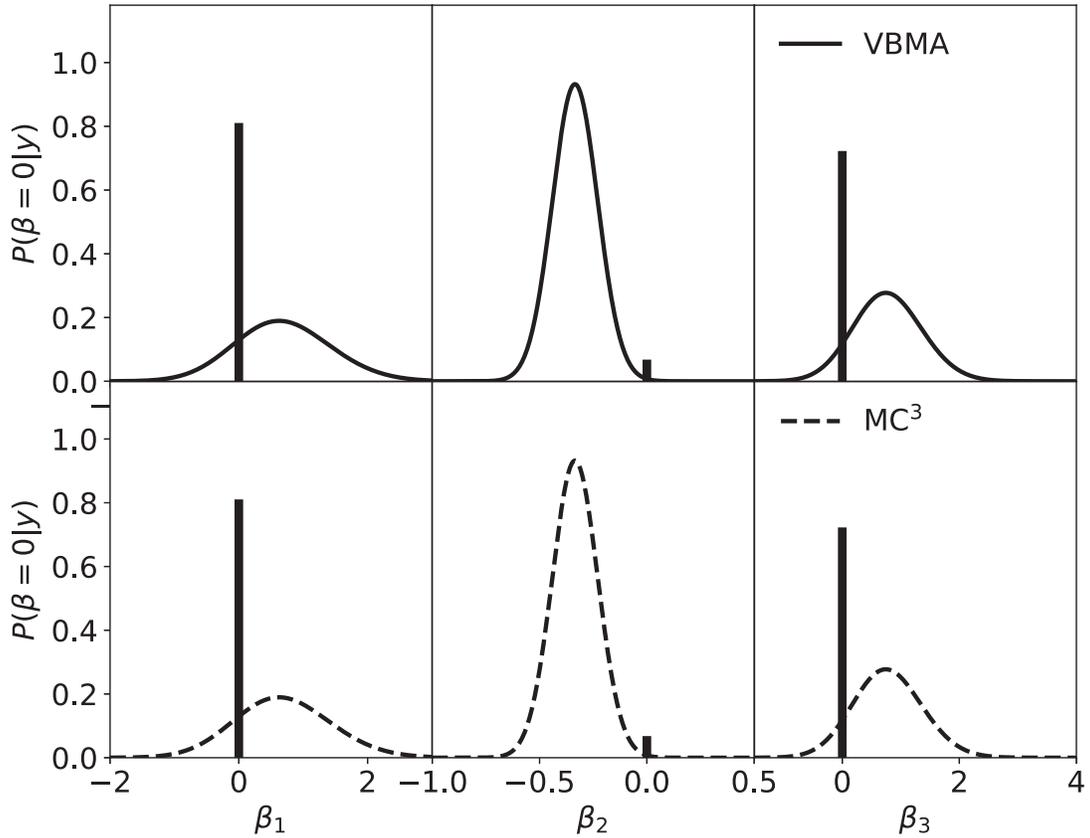}
	\end{center}
\end{figure}

Besides the model posterior probabilities, one can asses the fidelity of VBMA using the posterior distributions of regression coefficients based on the model average. Figure \ref{fig:post_betas_lin} shows the posterior distributions for the coefficients of the percentage of males 14-24, the probability of imprisonment, and the mean years of schooling based on the model averaging results. The figure additionally displays $\mathbb{P}(\beta = 0|\db)$ obtained by first summing the posterior model probabilities across the models for each predictor and then subtracting the value from one. We can see that the posterior distributions based on VBMA coincides with those obtained with MC$^3$. On the other hand, and somewhat expectantly, the computational overhead needed to compute the reparametrization gradients is unnecessarily large for the simple case of linear regression. The pre-training sequence took 4-10 seconds per model and the averaging of all 8 models took approximately 27 seconds on a mid-range laptop. Contrary to that, the MC$^3$ estimates were instantaneous for all the practical purposes. We shall start seeing the computational efficiency of VBMA in the subsequent applications.

\begin{figure}[h!]
\caption{Comparison between coverages of Bayesian predictive interval (equal-tails) on a testing set of 22 observations. The horizontal axis corresponds to the predictive coverage of MC$^3$ based intervals and the vertical axis to the coverage of VBMA. The diagonal line corresponds to the perfect agreement between the two methods.\label{fig:credibility_lin}}
	\begin{center}		
	\includegraphics[width=0.5\textwidth]{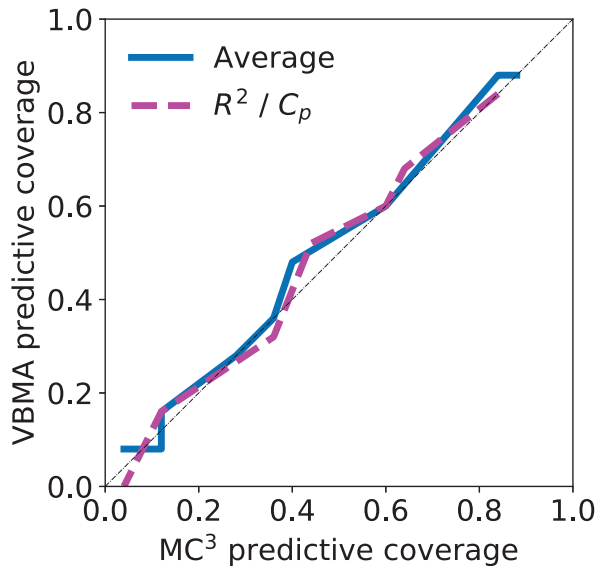}
	\end{center}
\end{figure}

\paragraph{Predictive Performance}
Similar to \cite{Raftery1997}, we asses the predictive ability of VBMA by randomly splitting the U.S. crime data into a training and a testing dataset. A 50-50 split was chosen here due to a relatively small size of the dataset. We subsequently re-run the VBMA (and MC$^3$) using the training dataset. The predictive performance was measured through coverage of Bayesian predictive intervals (equal-tails) with the credibility level ranging from $10\%$ to $90\%$ with $10\%$ increments. A $(1-\alpha)\times 100 \%$ prediction interval is a posterior credible interval within which a (predicted) observation falls with probability $(1-\alpha)$. An equal-tail interval is chosen so that the posterior probability of being below the interval is as likely as being above it \citep{BDA}. Figure \ref{fig:credibility_lin} shows the predictive coverage of the two methods plotted against each other with the diagonal dashed line indicating a perfect agreement between the methods. We can see that the coverages for the procedures match in general with small discrepancies at lower quantiles. Additionally, we compare the model averaging predictions with those obtained by the best models according to the adjusted $R^2$ and Mallows' $C_p$ under both VBMA and MC$^3$. Adjusted $R^2$ and Mallows' $C_p$ are commonly used model selection and evaluation criteria in a regression setting. Adjusted $R^2$ measures quality of the model in terms of total variability explained, whereas Mallows' $C_p$ estimates the size of the bias that is introduced into the predicted responses by having a model that is missing one or more important predictors \citep{James2013}. Both of these model selection strategies lead to $M_3$ as the best model. However, $M_3$ generally under-performed the model averaging and underestimated the declared coverage. We can see that by the general shift of the respected curve in comparison to the averaging results.

\subsection{Bayesian Logistic Regression}
Unlike the standard linear regression, generalized linear models such as logistic regression exemplify the slew of challenges that one can encounter when implementing BMA. First, the evaluation of the evidence integral does not have an analytic form and the integration can be high-dimensional. Additionally, direct sampling from the BMA posterior through MC$^3$ algorithm is not available. One therefore needs to approximate the evidence integral and consequently approximate the BMA posterior with MC samples from the mixture of the posteriors of each of the individual models.

Here we illustrate the utility of VBMA on the analysis of heart disease data \citep{Dua:2019} to asses the factors that contribute to the risk of heart attack. The models used are logistic regression models with logit link function of the form
\begin{equation}\label{eqn:logistic}
\log\bigg(\frac{\mathbb{P}(y=1)}{\mathbb{P}(y=0)}\bigg) = \beta_0 +  \sum_{j=1}^p \beta_j x_j,
\end{equation}
where $y=1$ corresponds to subjects with higher chance of heart attack, and $y=0$ to those with a smaller chance of heart attack. In this example, we shall consider five predictors, namely $x_1$ is the serum cholestoral in g/dl, $x_2$ is their resting blood pressure on admission to the hospital, $x_3$ is the biological sex, $x_4$ is the age, and $x_5$ is the maximum hear rate achieved during examination. This gives the total of 32 candidate models. All the continuous variables were again log transformed and centered. For the parameters in each model \eqref{eqn:logistic}, we use independent normal prior distributions.

\subsubsection{Results}
\begin{table}[h!]
\caption{\label{tab:log_prob} The top eight logistic regression models of the heart rate data according to their posterior model probabilities. The star indicates the inclusion of predictor in the model and the model ID is provided for easier referencing. Comparison between the VBMA and the MC based averaging is shown.}
\begin{center}
\begin{tabular}{|c|c|c|c|c|c|c|r|r|}
\cline{2-9}
\multicolumn{1}{l|}{} & \multicolumn{6}{c|}{\textbf{Inclusion}} & \multicolumn{2}{c|}{$p(M|\db)$} \\ \hline
\multicolumn{1}{|l|}{\textbf{Model}} & \multicolumn{1}{c|}{\textbf{Intercept}} & \multicolumn{1}{c|}{$\bm{x_1}$} & \multicolumn{1}{c|}{$\bm{x_2}$} & \multicolumn{1}{c|}{$\bm{x_3}$} & \multicolumn{1}{c|}{$\bm{x_4}$} & \multicolumn{1}{c|}{$\bm{x_5}$} & \multicolumn{1}{c|}{\textbf{MC}} & \multicolumn{1}{c|}{\textbf{VBMA}} \\ \hline
1 & * & * & * & *&  & * & 0.45  &  0.43  \\
2 & * & * & * & *& * & * &  0.28 & 0.28\\
3 & * & * &  & *&  * & * &  0.09 & 0.09 \\
4 & * & * &  & *&  & * &  0.06 & 0.06 \\
5 & * &  & * & * & * & * &  0.05  & 0.06 \\
6 & * &  & * & *&  & * & 0.04 & 0.05 \\
7 & * &  &  & *& * & * & 0.01 & 0.02\\
8 & * &  &  & *&  & * & $<0.01$ &  $<0.01$\\
\hline
\end{tabular}
\end{center}
\end{table}

Table \ref{tab:log_prob} shows the estimates of model posterior probabilities obtained with VBMA as compared to those computed using MC integration for the top 8 models. The VBMA results are based on a pre-training sequence of $500$ iterations with the model probabilities set to 1/32 and $100$ iterations of updating according to Algorithm \ref{alg:BBVBMA}. Ten MC samples from the variational distributions were used to estimate the ELBO gradient. The MC-based posterior model probabilities are based on $7.5 \times 10^5$ samples. This large number of samples was necessary in order to achieve reasonable convergence. We again observe a close match of the VBMA model posteriors with the MC model posteriors.

\begin{figure}[h!]
 \caption{Posterior distributions for predictor slopes based on VBMA (first row) and MCMC (second row). Namely, $\beta_1$ corresponds to the serum cholestoral in g/dl, $\beta_2$ to the resting blood pressure on admission to the hospital, $\beta_3$ to the biological sex, $\beta_4$ to the age, and $\beta_5$ is the maximum hear rate achieved during examination. The density is scaled so that the maximum of the density is equal to $\mathbb{P}(\beta \ne 0|\db)$. The spike corresponds to to $\mathbb{P}(\beta = 0|\db)$.\label{fig:betas_logistic}}
	\begin{center}		
	\includegraphics[width=1\textwidth]{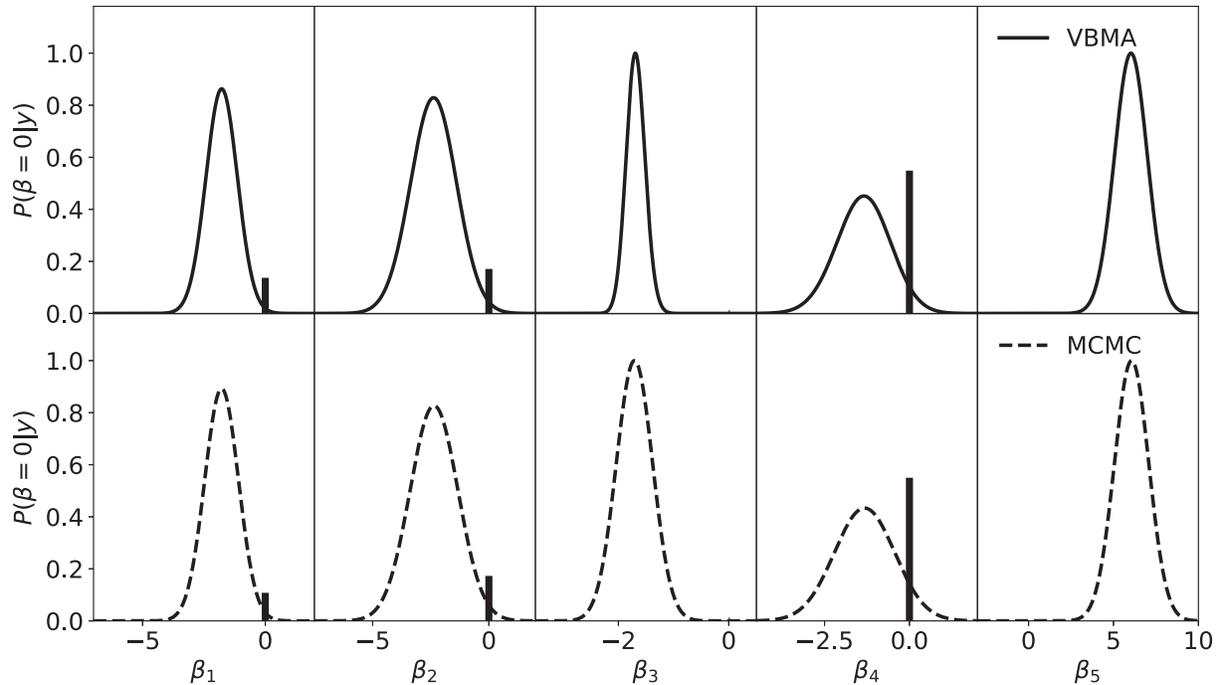}
	\end{center}
\end{figure}

Figure \ref{fig:betas_logistic} shows the posterior distribution of regression coefficients based on the model average. The MCMC results were obtained with No-U-Turn sampler \citep{NUTS} implemented in \texttt{Python} package for Bayesian statistical modeling \texttt{PyMC3} \citep{salvatier2016probabilistic}. Analogically to the linear regression example, VBMA algorithm with reparametrization gradients captures the posterior distributions well including the parameter uncertainties. When it comes to the computation times, we start seeing the benefits of VBMA for non-conjugate models. The pre-training sequence took 3-4 seconds per model and the averaging of all 32 models took approximately 20 seconds on a mid-range laptop. On the other hand, the MC estimates of evidence integrals required 20 minutes per model, and 20-50 seconds was needed to obtain $3\times 10^4$ samples via No-U-Turn sampler.

\subsection{Bayesian Model Selection}
Here, we demonstrate that the VBMA algorithm can be conveniently applied in the generalization of Bayesian hypotheses testing, that is, model selection with Bayes factors. Instead of averaging, suppose that we wish to compare the two Bayesian models,
\begin{equation*}
    M_0: \db \sim p(\db|\thetab_0), \thetab_0 \sim p(\thetab_0), \hspace{0.5cm} M_1: \db \sim p(\db|\thetab_1), \thetab_1 \sim p(\thetab_1),
\end{equation*}
where the definition of the parameter $\thetab$ may differ between models. Then, the Bayes factor $B_{01}$ in support of model $M_0$ is given by
\begin{equation}\label{eqn:BF}
B_{01} = \frac{p(\db|M_0)}{p(\db|M_1)}=\frac{p(M_0|\db)p(M_1)}{p(M_1|\db)p(M_0)},
\end{equation}
where $p(M_i|\db)$ for $i \in \{0,1\}$ is the model's posterior probability defined in equation \eqref{eqn:posteriorsMmodel:BMA}.
The quantity $B_{01}$ is the ratio of the posterior odds of model $M_0$ to its prior odds and represents the information about the evidence provided by the data in favor of model $M_0$ as opposed to $M_1$ \citep{KassRaftery1995}. It should be clear from the definition of \eqref{eqn:BF} that the Bayesian model selection suffers from exactly the same computational challenges as BMA. To this extent, VBMA directly approximates posterior probabilities of individual models and Bayes factors can be conveniently computed as a byproduct of the algorithm without the need of approximating the model evidence \eqref{eqn:evidencel:BMA}.

\subsubsection{Linear and Logistic Regression Examples}

To illustrate the Bayesian model selection via VBMA, we consider the following hypotheses for both linear and logistic regression examples above and compare the VBMA based results with their MC counterparts:
\begin{equation}\label{eqn:hypothese}
    H_0: \beta_1 = 0, \hspace{0.5cm} H_1: \beta_1 \ne 0.
\end{equation}

For the linear regression case, this corresponds to comparing models $M_1$ and $M_3$. For the logistic regression example, we need to compare models $M_5$ and $M_2$. Table \ref{tab:bayes_factors} presents the respective Bayes factor approximations. For both linear and logistic regression examples, VBMA approximations qualitatively agree with the MC based approximations. The results show that the U.S crime data favor the linear regression model with $\beta_1 = 0$, and the heart disease data favor logistic regression model with $\beta_1 \ne 0$.

\begin{table}[h!]
\caption{\label{tab:bayes_factors} Bayes factors obtained via VBMA approximation and MC methods. Models $M_1$ and $M_3$ are considered for the linear regression example and models $M_5$ and $M_2$ for the logistic regression example. Bayes factor larger than 1 indicates selection of the model with $\beta_1 = 0$ and vice versa.}
\begin{center}
\begin{tabular}{l|r|r|}
\cline{2-3}
 & \multicolumn{2}{c|}{\textbf{Bayes Factor}} \\ \hline
\multicolumn{1}{|l|}{\textbf{Example}} & \multicolumn{1}{l|}{\textbf{MC}} & \multicolumn{1}{l|}{\textbf{VBMA}} \\ \hline
\multicolumn{1}{|l|}{Linear regression} & 2.43 & 3.00 \\ 
\multicolumn{1}{|l|}{Logistic regression} & 0.18& 0.21 \\ \hline
\end{tabular}
\end{center}
\end{table}

\subsection{Nuclear Mass Predictions}
As an illustration of VBMA in a scenario where application of standard MCMC-based inference is challenging in practice, we study the separation energies of atomic nuclei which were the subject of various recent machine learning applications (Gaussian process modeling) in the field of nuclear physics \citep{Neufcourt2018,Neufcourt2019,Neufcourt2020a,Neufcourt2020b}. Namely, our focus is the two-neutron separation energy ($S_{2n}$) which is a fundamental property of atomic nucleus and is defined as the energy required to remove two neutrons from the nucleus. The $S_{2n}$ values can be obtained through a nuclear mass difference. The knowledge of separation energies determines the limits of nuclear existence and predictions of these quantities can help guide the experimental research at future rare isotope facilities.

In this example, we shall consider 6 state-of-the-art nuclear mass models based on the nuclear density functional theory (DFT) \citep{Nazarewicz2016}: the Skyrme energy density functionals SkM$^*$ \citep{SKMstar}, SkP \citep{SKP}, SLy4 \citep{SLY4}, SV-min \citep{SVMIN}, UNEDF0 \cite{UNEDF0}, and UNEDF1 \citep{UNEDF1}. These are global nuclear mass models, because the are capable of reliably describing the whole nuclear chart. Our analysis closely follows that of \cite{Neufcourt2020b}, where we consider the statistical model for the differences $y_i = S_{2n}^{exp}(\xb_i) - S_{2n}^{th}(\xb_i)$ between the observed experimental data and the predictions given by the theoretical models of the form
\begin{equation}\label{}
    y_i = f(\xb_i) + \sigma \epsilon_i,
\end{equation}
where $\xb_i = (Z_i, N_i)$ corresponds to the the proton number $Z_i$ and the neutron number $N_i$ of a nucleus. The function $f(\cdot)$ represents the systematic discrepancy between the underlying physical process and the theoretical mass model. The quantity $\sigma \epsilon_i$ is the scaled experimental error which is assumed to be i.i.d. normal with mean zero. For the systematic discrepancy, \cite{Neufcourt2020b} take a Gaussian process (GP) on the two dimensional space $x = (Z,N)$:
\begin{equation}\label{eqn:GP_model}
    f(\xb) \sim \mathcal{GP}(\beta, k(\xb, \xb')),
\end{equation}
where $\beta$ is the constant mean and $k$ is the squared exponential covariance function characterized by the scale $\eta$ and characteristic correlation ranges $\nu_Z$ and $\nu_N$:
\begin{align}\label{eqn:gp_ker}
    k(\xb, \xb') = \eta^2 e^{-\frac{(Z-Z')^2}{2\nu_Z^2} -\frac{(N-N')^2}{2\nu_N^2}}.
\end{align}
The GP with covariance \eqref{eqn:gp_ker} is a sensible nonparametric model for the systematic discrepancy as it is expected to be relatively smooth and stationary \citep{Neufcourt2020}. Since neither of the six Skyrme energy functionals a-priory stands out on the full nuclear domain, using BMA for averaging or model selection is a logical approach here that will allow for predictions with realistically quantified uncertainties. 

\begin{figure}[h!]
 \caption{The nuclear chart of even-even and odd-even nuclei divided into the training and testing datasets for the GP modeling of the residuals of two-neutron separation energies $S_{2n}$. $Z$ corresponds to the proton number and $N$ is the neutron number. \label{fig:domains}}
	\begin{center}		
	\includegraphics[width=0.7\textwidth]{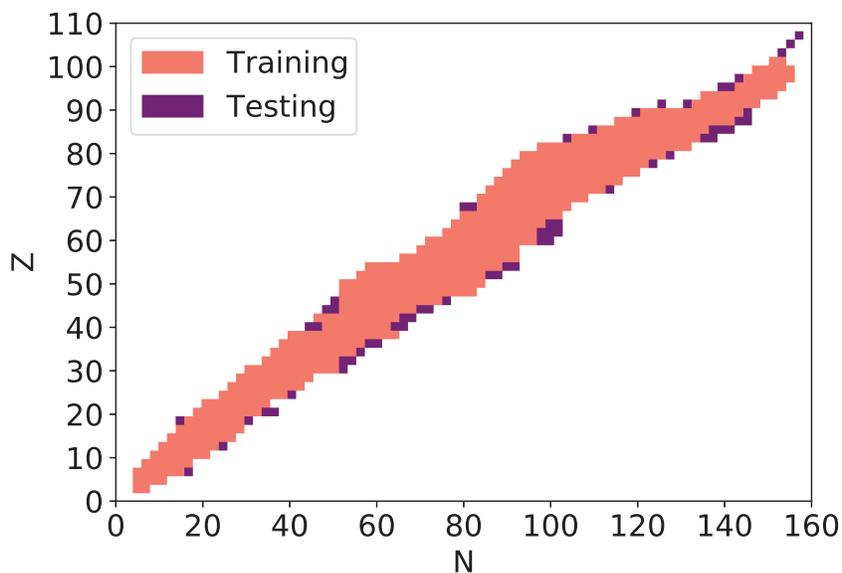}
	\end{center}
\end{figure}

As the experimental observations, we take the most recent measured values of two-neutron separation energies from the AME2003 dataset \citep{AME2003} as training data ($n = 1029$)
and keep all additional data tabulated in AME2016 \citep{AME2016} for a testing dataset $(n=120)$. The domains of these datasets are depicted in Figure \ref{fig:domains}. Note that we use both even-even (meaning both $Z$ and $N$ are even) and odd-even nuclei jointly for the training to fully account for the correlations between systematic discrepancies unlike \cite{Neufcourt2020b} who fitted independent GPs on the two domains separately to make the computations manageable. Using the proposed VBMA approach, we are able to do computations in matter of minutes which would take dozens of hours using the standard MCMC approximation.

\subsection{Results}
The performance of VBMA in averaging of the GP enhanced nuclear mass models was compared with BMA based on the posterior approximation by No-U-Turn sampler and the MC estimates of evidence integrals. Similarly to the previous examples, the VBMA results are based on a pre-training sequence of $300$ iterations with the model probabilities set to 1/6. This pre-training sequence lead to the selection of UNEDF1 ($p(M|\db) = 1$) with stable ELBOs and so no further training was needed. Ten MC samples from the variational distributions were used to estimate the reparametrization gradient gradient. The resulting \textit{root-mean-square error} (RMSE) on the testing dataset of 120 nuclei was 0.406 MeV as compared to the RMSE of 0.419 MeV given by the MCMC approximation. These RMSE values are consistent with those obtained by \cite{Neufcourt2020b}. The MCMC results are based on $2 \times 10^4$ posterior samples ($10^4$ burn-in) and the MC posterior model probabilities are based on $2 \times 10^5$ samples which also lead to the selection of UNEDF1.

The MCMC implementation proved to be significantly more time consuming. It required 15-20 hours per model to generate the posterior samples and about six hours for  the MC integration. On the other hand, the variational approach required between 20 to 25 minutes per model which clearly demonstrates the utility of VBMA in more complex modeling scenarios. The fidelity of Bayesian predictive intervals was equivalent between the VBMA and MCMC similarly to the linear regression example. We refer the reader to the Supplement for additional results containing the study of predictive coverage.

\section{Discussion}

We presented a VBI approach to BMA that avoids some of the practical challenges that burdens the standard MCMC based approaches to approximate the BMA posterior, especially numerical evaluation of the evidence integral and long sampling times of MCMC sampler. The fidelity of the method was demonstrated on a series of pedagogical examples including the averaging of linear and logistic regression models and Bayesian model selection via Bayes factors. To fully showcase the computational benefits of VBMA, we applied our methodology to nuclear mass models with GP model for systematic discrepancies. The observed speed-up in the case of GP modeling was at least fifty-fold compared to the standard MCMC approaches.

The proposed procedure is ``black box'' in the sense that it can be readily applied to wide range of models with minimal additional derivations needed. For instance, VBMA can be conveniently applied to nonconjugate models including generalized linear models, Bayesian neural networks, and Deep latent Gaussian models \citep{VarInfTut}.

Additionally, VBMA is a general VBI algorithm and the presented implementation with the Adam learning rate and the mean-field variational family is just one of many implementations. One can consider any adaptive learning rate and other variational family available in the literature. VBMA can also be simply modified for greater scalability in the scenarios with complex machine learning models that need to be fitted to massive datasets. First, one can subsample from the data to construct computationally cheap noisy estimates of ELBO gradients. Secondly, the nature of VBMA allows for immediate parallelization  across the models. To achieve a faster convergence of the algorithm, VBMA can be augmented with Rao-Blackwellizaiton \citep{RB}, control variates \citep{Ross:2006}, and importance sampling \cite{OBBVI} to even further reduce the variance of noisy gradient estimators. 

Of course, the use of VBI comes at a cost and one cannot avoid the general pitfalls of variational methods. Using mean-field families can lead to posterior distributions with underestimated uncertainties in cases of highly correlated parameters. One can improve the fidelity of posteriors by using more complex variational family that does not assume independence of unknown parameters \citep{VarInfTut, VIth}. The choice of adaptive learning rate can be sometimes challenging in practice, and one may observe significant differences among the adaptive learning rates, and a careful sensitivity analysis must be performed. For the models considered in this work, we do not observe significant differences between the RMSprop and the Adam (see the Supplement). The reparametrization gradient used in Algorithm \ref{alg:BBVBMA} works only for differentiable models with no discrete variables. One can use the score gradient \citep{Ranganath14} for models with discrete variables which is also black box in nature, however, the variance of the gradient estimates is larger than that of reparametrization gradients. Lastly, the computational overhead of VBMA for simple models (such as linear regression) can be too high to achieve any meaningful advantage, however, the use of VBMA in complex models can lead to significant computational gains.

\subsection*{Acknowledgments}

The authors thank the reviewers, the Associate Editor, and the Editor for their helpful comments and ideas. This work was supported in part through computational resources and services provided by the Institute for Cyber-Enabled Research at Michigan State University.

\subsection*{Funding}
The research is partially supported by the National Science Foundation funding DMS-1952856 and OAC-2004601.

\bigskip
\begin{appendices}
\section{Parametrization of Variational Families} \label{sec:appendix:reparametrization}
\subsection{Normal Variational Family}
Let us consider a real valued parameter $\theta$ with normal variation family $q(\theta|\mu, \sigma^2)$ parametrized by the mean $\mu$ and variance $\sigma^2$. Under the transformation \eqref{eqn:reparametrization}, we get the following expressions for the log likelihood of the variational distribution
\begin{equation}
    \log q(\theta| \mu, \lambda_{\sigma}) = - \frac{1}{2} \log [ \log(e^{\lambda_{\sigma}} + 1)] - \frac{1}{2} \log 2 \pi - \frac{1}{2} \frac{(\theta - \mu)^2}{\log(e^{\lambda_{\sigma}} + 1)}.
\end{equation}
The reparametrization gradient is then obtained with $z\sim \text{Normal}(0,1)$ so that $$\theta = t(z, (\mu, \lambda_{\sigma})) = z \times \sqrt{\log(e^{\lambda_{\sigma}} + 1)} + \mu.$$

\subsection{Log-normal Variational Family}
For a positive-valued parameters, we shall consider a log-normal variational family $q(\theta | m, \tau^2)$ parametrized by the mean $m$ and variance $\tau ^ 2$:
\begin{equation*}
    \log q(\theta | m, \lambda_\tau) = - \log\theta -\frac{1}{2} \log [\log(e^{\lambda_{\tau}} + 1)] - \frac{1}{2} \log 2 \pi - \frac{1}{2} \frac{(\log \theta - \mu)^2}{\log(e^{\lambda_{\tau}} + 1)}
\end{equation*}
The reparametrization gradient is then obtained with $z\sim \text{Normal}(0,1)$ so that $$\theta = t(z, (\mu, \lambda_{\tau})) = e^{z \times \sqrt{\log(e^{\lambda_{\tau}} + 1)} + \mu}.$$
\end{appendices}

\bibliographystyle{Chicago}

\bibliography{biblio}

\begin{thebibliography}{}

\bibitem[\protect\citeauthoryear{Ambrogioni, Lin, Fertig, Vikram, Hinne, Moore,
  and van Gerven}{Ambrogioni et~al.}{2021}]{pmlr-v130-ambrogioni21a}
Ambrogioni, L., K.~Lin, E.~Fertig, S.~Vikram, M.~Hinne, D.~Moore, and M.~van
  Gerven (2021, 13--15 Apr).
\newblock Automatic structured variational inference.
\newblock In A.~Banerjee and K.~Fukumizu (Eds.), {\em Proceedings of The 24th
  International Conference on Artificial Intelligence and Statistics}, Volume
  130 of {\em Proceedings of Machine Learning Research}, pp.\  676--684. PMLR.

\bibitem[\protect\citeauthoryear{Ardia, Baştürk, Hoogerheide, and {van
  Dijk}}{Ardia et~al.}{2012}]{ARDIA2012}
Ardia, D., N.~Baştürk, L.~Hoogerheide, and H.~K. {van Dijk} (2012).
\newblock A comparative study of monte carlo methods for efficient evaluation
  of marginal likelihood.
\newblock {\em Computational Statistics and Data Analysis\/}~{\em 56\/}(11),
  3398--3414.
\newblock 1st issue of the Annals of Computational and Financial Econometrics
  Sixth Special Issue on Computational Econometrics.

\bibitem[\protect\citeauthoryear{Audi, Wapstra, and Thibault}{Audi
  et~al.}{2003}]{AME2003}
Audi, G., A.~Wapstra, and C.~Thibault (2003).
\newblock The {{AME}}2003 atomic mass evaluation: (ii). tables, graphs and
  references.
\newblock {\em Nuclear Physics A\/}~{\em 729}, 337--676.

\bibitem[\protect\citeauthoryear{Balasubramanian, Visweswaran, Cooper, and
  Gopalakrishnan}{Balasubramanian et~al.}{2014}]{med-Vi14}
Balasubramanian, J.~B., S.~Visweswaran, G.~F. Cooper, and V.~Gopalakrishnan
  (2014).
\newblock Selective model averaging with {{Bayesian}} rule learning for
  predictive biomedicine.
\newblock {\em AMIA Joint Summits on Translational Science proceedings AMIA
  Summit on Translational Science\/}~{\em 2014}, 17--22.

\bibitem[\protect\citeauthoryear{Bartel, Quentin, Brack, Guet, and
  H{\aa}kansson}{Bartel et~al.}{1982}]{SKMstar}
Bartel, J., P.~Quentin, M.~Brack, C.~Guet, and H.-B. H{\aa}kansson (1982).
\newblock Towards a better parametrisation of {Skyrme-like} effective forces: a
  critical study of the {SkM} force.
\newblock {\em Nuclear Physics A\/}~{\em 386\/}(1), 79--100.

\bibitem[\protect\citeauthoryear{Bernardo and Smith}{Bernardo and
  Smith}{1994}]{Bernardo1994}
Bernardo, J.~M. and A.~F.~M. Smith (1994).
\newblock {\em Reference analysis}, Chapter Inference.
\newblock Wiley.

\bibitem[\protect\citeauthoryear{Bhattacharya and Maiti}{Bhattacharya and
  Maiti}{2021}]{BHATTACHARYA2021151}
Bhattacharya, S. and T.~Maiti (2021).
\newblock Statistical foundation of variational bayes neural networks.
\newblock {\em Neural Networks\/}~{\em 137}, 151--173.

\bibitem[\protect\citeauthoryear{Blei, Kucukelbir, and McAuliffe}{Blei
  et~al.}{2017}]{VarInfTut}
Blei, D.~M., A.~Kucukelbir, and J.~D. McAuliffe (2017).
\newblock Variational inference: A review for statisticians.
\newblock {\em Journal of the American Statistical Association\/}~{\em
  112\/}(518), 859--877.

\bibitem[\protect\citeauthoryear{Bottou, {Le Cun}, and Bengio}{Bottou
  et~al.}{1997}]{bottou-97}
Bottou, L., Y.~{Le Cun}, and Y.~Bengio (1997).
\newblock Global training of document processing systems using graph
  transformer networks.
\newblock In {\em Proceedings of Computer Vision and Pattern Recognition
  (CVPR)}, pp.\  489--493. IEEE.

\bibitem[\protect\citeauthoryear{Casella and Robert}{Casella and
  Robert}{1996}]{RB}
Casella, G. and C.~P. Robert (1996).
\newblock Rao-blackwellisation of sampling schemes.
\newblock {\em Biometrika\/}~{\em 83\/}(1), 81--94.

\bibitem[\protect\citeauthoryear{Chabanat, Bonche, Haensel, Meyer, and
  Schaeffer}{Chabanat et~al.}{1995}]{SLY4}
Chabanat, E., P.~Bonche, P.~Haensel, J.~Meyer, and R.~Schaeffer (1995).
\newblock New {{Skyrme}} effective forces for supernovae and neutron rich
  nuclei.
\newblock {\em Physica Scr.\/}~{\em 1995\/}(T56), 231.

\bibitem[\protect\citeauthoryear{Clarke}{Clarke}{2003}]{Clarke2003}
Clarke, B. (2003, 11).
\newblock Comparing bayes model averaging and stacking when model approximation
  error cannot be ignored.
\newblock {\em The Journal of Machine Learning Research\/}~{\em 4}, 683--712.

\bibitem[\protect\citeauthoryear{Clyde and Iversen}{Clyde and
  Iversen}{2013}]{Iversen2013}
Clyde, M. and E.~Iversen (2013, 01).
\newblock {\em Bayesian Model Averaging in the M-Open Framework}, pp.\
  484--498.

\bibitem[\protect\citeauthoryear{Clyde, Ghosh, and Littman}{Clyde
  et~al.}{2011}]{ML-Mer11}
Clyde, M.~A., J.~Ghosh, and M.~L. Littman (2011).
\newblock Bayesian adaptive sampling for variable selection and model
  averaging.
\newblock {\em Journal of Computational and Graphical Statistics\/}~{\em 20},
  80--101.

\bibitem[\protect\citeauthoryear{Dobaczewski, Flocard, and Treiner}{Dobaczewski
  et~al.}{1984}]{SKP}
Dobaczewski, J., H.~Flocard, and J.~Treiner (1984).
\newblock {Hartree-Fock-Bogolyubov} description of nuclei near the neutron-drip
  line.
\newblock {\em Nucl. Phys. A\/}~{\em 422\/}(1), 103--139.

\bibitem[\protect\citeauthoryear{Dua and Graff}{Dua and Graff}{2017}]{Dua:2019}
Dua, D. and C.~Graff (2017).
\newblock {UCI} machine learning repository.

\bibitem[\protect\citeauthoryear{Duchi, Hazan, and Singer}{Duchi
  et~al.}{2011}]{AdaGrad}
Duchi, J., E.~Hazan, and Y.~Singer (2011).
\newblock Adaptive subgradient methods for online learning and stochastic
  optimization.
\newblock {\em The Journal of Machine Learning Research\/}~{\em 12},
  2121--2159.

\bibitem[\protect\citeauthoryear{Feroz, Hobson, and Bridges}{Feroz
  et~al.}{2009}]{MultiNest}
Feroz, F., M.~P. Hobson, and M.~Bridges (2009).
\newblock Multinest: an efficient and robust bayesian inference tool for
  cosmology and particle physics.
\newblock {\em Monthly Notices of the Royal Astronomical Society\/}~{\em 398},
  1601–1614.

\bibitem[\protect\citeauthoryear{Fortunato, Blundell, and Vinyals}{Fortunato
  et~al.}{2017}]{fortunato}
Fortunato, M., C.~Blundell, and O.~Vinyals (2017).
\newblock Bayesian recurrent neural networks.
\newblock {\em arXiv preprint arXiv: 1704.02798\/}.

\bibitem[\protect\citeauthoryear{Fragoso, Bertoli, and Louzada}{Fragoso
  et~al.}{2018}]{Fragoso}
Fragoso, T.~M., W.~Bertoli, and F.~Louzada (2018).
\newblock Bayesian model averaging: A systematic review and conceptual
  classification.
\newblock {\em International Statistical Review\/}~{\em 86}, 1--28.

\bibitem[\protect\citeauthoryear{Friel and Pettitt}{Friel and
  Pettitt}{2008}]{Freil2008}
Friel, N. and A.~N. Pettitt (2008).
\newblock Marginal likelihood estimation via power posteriors.
\newblock {\em Journal of the Royal Statistical Society: Series B (Statistical
  Methodology)\/}~{\em 70\/}(3), 589--607.

\bibitem[\protect\citeauthoryear{Friel and Wyse}{Friel and
  Wyse}{2012}]{Friel2012}
Friel, N. and J.~Wyse (2012).
\newblock Estimating the evidence – a review.
\newblock {\em Statistica Neerlandica\/}~{\em 66\/}(3), 288--308.

\bibitem[\protect\citeauthoryear{Gelman, Carlin, Stern, Dunson, Vehtari, and
  Rubin}{Gelman et~al.}{2013}]{BDA}
Gelman, A., J.~Carlin, H.~Stern, D.~Dunson, A.~Vehtari, and D.~Rubin (2013).
\newblock {\em Bayesian Data Analysis\/} (Third ed.).
\newblock CRC Pres.

\bibitem[\protect\citeauthoryear{Geweke}{Geweke}{1999}]{Geweke}
Geweke, J. (1999).
\newblock Using simulation methods for {{Bayesian}} econometric models:
  inference, development, and communication.
\newblock {\em Econometric Reviews\/}~{\em 18}, 1--73.

\bibitem[\protect\citeauthoryear{Hern{\'a}ndez, Raftery, Pennington, and
  Parnell}{Hern{\'a}ndez et~al.}{2018}]{ML-Hernandez2018}
Hern{\'a}ndez, B., A.~E. Raftery, S.~R. Pennington, and A.~C. Parnell (2018).
\newblock Bayesian additive regression trees using {{Bayesian}} model
  averaging.
\newblock {\em Statistics and Computing\/}~{\em 28}, 869--890.

\bibitem[\protect\citeauthoryear{Hoeting, Madigan, Raftery, and
  Volinsky}{Hoeting et~al.}{1999}]{BMA}
Hoeting, J.~A., D.~Madigan, A.~E. Raftery, and C.~T. Volinsky (1999).
\newblock Bayesian model averaging: A tutorial.
\newblock {\em Statistical Science\/}~{\em 14}, 382--401.

\bibitem[\protect\citeauthoryear{Hoffman and Blei}{Hoffman and
  Blei}{2015}]{pmlr-v38-hoffman15}
Hoffman, M. and D.~Blei (2015, 09--12 May).
\newblock {Stochastic structured variational inference}.
\newblock In {\em Proceedings of the Eighteenth International Conference on
  Artificial Intelligence and Statistics}, Volume~38, San Diego, CA, pp.\
  361--369. PMLR.

\bibitem[\protect\citeauthoryear{Homan and Gelman}{Homan and
  Gelman}{2014}]{NUTS}
Homan, M.~D. and A.~Gelman (2014).
\newblock The {{No-U-Turn Sampler}}: Adaptively setting path lengths in
  {{Hamiltonian Monte Carlo}}.
\newblock {\em Journal of Machine Learning Research\/}~{\em 15}, 1351--1381.

\bibitem[\protect\citeauthoryear{Hooten and Hobbs}{Hooten and
  Hobbs}{2015}]{ecol-Ho15}
Hooten, M.~B. and N.~T. Hobbs (2015).
\newblock A guide to {{Bayesian}} model selection for ecologists.
\newblock {\em Ecological Monographs\/}~{\em 85}, 3--28.

\bibitem[\protect\citeauthoryear{James, Witten, Hastie, and Tibshirani}{James
  et~al.}{2013}]{James2013}
James, G., D.~Witten, T.~Hastie, and R.~Tibshirani (2013).
\newblock {\em An Introduction to Statistical Learning: with Applications in
  R}.
\newblock New York, NY: Springer New York.

\bibitem[\protect\citeauthoryear{Jaureguiberry, Vincent, and
  Richard}{Jaureguiberry et~al.}{2014}]{VBMA2014}
Jaureguiberry, X., E.~Vincent, and G.~Richard (2014).
\newblock Variational bayesian model averaging for audio source separation.
\newblock In {\em 2014 IEEE Workshop on Statistical Signal Processing (SSP)},
  pp.\  33--36.

\bibitem[\protect\citeauthoryear{Jordan, Ghahramani, Jaakkola, and Saul}{Jordan
  et~al.}{1999}]{Jordan1999}
Jordan, M.~I., Z.~Ghahramani, T.~S. Jaakkola, and L.~K. Saul (1999).
\newblock An introduction to variational methods for graphical models.
\newblock {\em Machine Learning\/}~{\em 37}, 183--233.

\bibitem[\protect\citeauthoryear{Kass and Raftery}{Kass and
  Raftery}{1995}]{KassRaftery1995}
Kass, R.~E. and A.~E. Raftery (1995).
\newblock Bayes factors.
\newblock {\em Journal of the American Statistical Association\/}~{\em 90},
  773--795.

\bibitem[\protect\citeauthoryear{Kejzlar, Neufcourt, Nazarewicz, and
  Reinhard}{Kejzlar et~al.}{2020}]{Kejzlar2020}
Kejzlar, V., L.~Neufcourt, W.~Nazarewicz, and P.-G. Reinhard (2020, jul).
\newblock Statistical aspects of nuclear mass models.
\newblock {\em Journal of Physics G: Nuclear and Particle Physics\/}~{\em
  47\/}(9), 094001.

\bibitem[\protect\citeauthoryear{Kingma and Ba}{Kingma and Ba}{2014}]{Adam}
Kingma, D. and J.~Ba (2014, 12).
\newblock Adam: A method for stochastic optimization.
\newblock {\em International Conference on Learning Representations\/}.

\bibitem[\protect\citeauthoryear{Kingma, Salimans, Jozefowicz, Chen, Sutskever,
  and Welling}{Kingma et~al.}{2016}]{NIPS2016_ddeebdee}
Kingma, D.~P., T.~Salimans, R.~Jozefowicz, X.~Chen, I.~Sutskever, and
  M.~Welling (2016).
\newblock Improved variational inference with inverse autoregressive flow.
\newblock In D.~Lee, M.~Sugiyama, U.~Luxburg, I.~Guyon, and R.~Garnett (Eds.),
  {\em Advances in Neural Information Processing Systems}, Volume~29. Curran
  Associates, Inc.

\bibitem[\protect\citeauthoryear{Kl{\"{u}}pfel, Reinhard, B{\"{u}}rvenich, and
  Maruhn}{Kl{\"{u}}pfel et~al.}{2009}]{SVMIN}
Kl{\"{u}}pfel, P., P.-G. Reinhard, T.~J. B{\"{u}}rvenich, and J.~A. Maruhn
  (2009, Mar).
\newblock Variations on a theme by {Skyrme}: A systematic study of adjustments
  of model parameters.
\newblock {\em Phys. Rev. C\/}~{\em 79\/}(3), 034310.

\bibitem[\protect\citeauthoryear{Kobyzev, Prince, and Brubaker}{Kobyzev
  et~al.}{2021}]{normalizing}
Kobyzev, I., S.~J. Prince, and M.~A. Brubaker (2021, Nov).
\newblock Normalizing flows: An introduction and review of current methods.
\newblock {\em IEEE Transactions on Pattern Analysis and Machine
  Intelligence\/}~{\em 43\/}(11), 3964–3979.

\bibitem[\protect\citeauthoryear{Kortelainen, Lesinski, Mor{\'e}, Nazarewicz,
  Sarich, Schunck, Stoitsov, and Wild}{Kortelainen et~al.}{2010}]{UNEDF0}
Kortelainen, M., T.~Lesinski, J.~J. Mor{\'e}, W.~Nazarewicz, J.~Sarich,
  N.~Schunck, M.~V. Stoitsov, and S.~M. Wild (2010).
\newblock Nuclear energy density optimization.
\newblock {\em Physical Review C\/}~{\em 82\/}(2), 024313.

\bibitem[\protect\citeauthoryear{Kortelainen, McDonnell, Nazarewicz, Reinhard,
  Sarich, Schunck, Stoitsov, and Wild}{Kortelainen et~al.}{2012}]{UNEDF1}
Kortelainen, M., J.~McDonnell, W.~Nazarewicz, P.-G. Reinhard, J.~Sarich,
  N.~Schunck, M.~V. Stoitsov, and S.~M. Wild (2012).
\newblock Nuclear energy density optimization: large deformations.
\newblock {\em Physical Review C\/}~{\em 85}, 024304.

\bibitem[\protect\citeauthoryear{Kucukelbir, Tran, Ranganath, Gelman, and
  Blei}{Kucukelbir et~al.}{2017}]{Kucukelbir2017}
Kucukelbir, A., D.~Tran, R.~Ranganath, A.~Gelman, and D.~M. Blei (2017).
\newblock Automatic differentiation variational inference.
\newblock {\em Journal of Machine Learning Research\/}~{\em 18\/}(14), 1--45.

\bibitem[\protect\citeauthoryear{Latouche and Robin}{Latouche and
  Robin}{2016}]{latoucheVBMA}
Latouche, P. and S.~S. Robin (2016).
\newblock {Variational Bayes model averaging for graphon functions and motif
  frequencies inference in W-graph models}.
\newblock {\em {Statistics and Computing}\/}~{\em 26\/}(6), 1173 -- 1185.

\bibitem[\protect\citeauthoryear{Leamer}{Leamer}{1978}]{Leamer1978}
Leamer, E.~E. (1978).
\newblock {\em Specification Searches: Ad Hoc Inference with Nonexperimental
  Data}.
\newblock Wiley.

\bibitem[\protect\citeauthoryear{Lenk}{Lenk}{2009}]{Lenk2009}
Lenk, P. (2009).
\newblock Simulation pseudo-bias correction to the harmonic mean estimator of
  integrated likelihoods.
\newblock {\em Journal of Computational and Graphical Statistics\/}~{\em
  18\/}(4), 941--960.

\bibitem[\protect\citeauthoryear{Madigan, Gavrin, and Raftery}{Madigan
  et~al.}{1995}]{Madigan1995}
Madigan, D., J.~Gavrin, and A.~E. Raftery (1995).
\newblock Eliciting prior information to enhance the predictive performance of
  bayesian graphical models.
\newblock {\em Communications in Statistics - Theory and Methods\/}~{\em
  24\/}(9), 2271--2292.

\bibitem[\protect\citeauthoryear{Masegosa}{Masegosa}{2020}]{Masegosa2020}
Masegosa, A. (2020).
\newblock Learning under model misspecification: Applications to variational
  and ensemble methods.
\newblock In H.~Larochelle, M.~Ranzato, R.~Hadsell, M.~F. Balcan, and H.~Lin
  (Eds.), {\em Advances in Neural Information Processing Systems}, Volume~33,
  pp.\  5479--5491. Curran Associates, Inc.

\bibitem[\protect\citeauthoryear{Mukhopadhyay and Dunson}{Mukhopadhyay and
  Dunson}{2020}]{Dunson2020}
Mukhopadhyay, M. and D.~B. Dunson (2020).
\newblock Targeted random projection for prediction from high-dimensional
  features.
\newblock {\em Journal of the American Statistical Association\/}~{\em
  115\/}(532), 1998--2010.

\bibitem[\protect\citeauthoryear{Nazarewicz}{Nazarewicz}{2016}]{Nazarewicz2016}
Nazarewicz, W. (2016).
\newblock Challenges in nuclear structure theory.
\newblock {\em Journal of Physics G: Nuclear and Particle Physics\/}~{\em 43},
  044002.

\bibitem[\protect\citeauthoryear{Neal}{Neal}{2001}]{Neal2001}
Neal, R. (2001, 01).
\newblock Annealed importance sampling.
\newblock {\em Statistics and Computing\/}~{\em 11}.

\bibitem[\protect\citeauthoryear{Neufcourt, Cao, Giuliani, Nazarewicz, Olsen,
  and Tarasov}{Neufcourt et~al.}{2020a}]{Neufcourt2020}
Neufcourt, L., Y.~Cao, S.~Giuliani, W.~Nazarewicz, E.~Olsen, and O.~B. Tarasov
  (2020a, Jan).
\newblock Beyond the proton drip line: Bayesian analysis of proton-emitting
  nuclei.
\newblock {\em Physical Review C\/}~{\em 101}, 014319.

\bibitem[\protect\citeauthoryear{Neufcourt, Cao, Giuliani, Nazarewicz, Olsen,
  and Tarasov}{Neufcourt et~al.}{2020b}]{Neufcourt2020a}
Neufcourt, L., Y.~Cao, S.~Giuliani, W.~Nazarewicz, E.~Olsen, and O.~B. Tarasov
  (2020b, Jan).
\newblock Beyond the proton drip line: Bayesian analysis of proton-emitting
  nuclei.
\newblock {\em Phys. Rev. C\/}~{\em 101}, 014319.

\bibitem[\protect\citeauthoryear{Neufcourt, Cao, Giuliani, Nazarewicz, Olsen,
  and Tarasov}{Neufcourt et~al.}{2020c}]{Neufcourt2020b}
Neufcourt, L., Y.~Cao, S.~A. Giuliani, W.~Nazarewicz, E.~Olsen, and O.~B.
  Tarasov (2020c, Apr).
\newblock Quantified limits of the nuclear landscape.
\newblock {\em Phys. Rev. C\/}~{\em 101}, 044307.

\bibitem[\protect\citeauthoryear{Neufcourt, Cao, Nazarewicz, Olsen, and
  Viens}{Neufcourt et~al.}{2019}]{Neufcourt2019}
Neufcourt, L., Y.~Cao, W.~Nazarewicz, E.~Olsen, and F.~Viens (2019, Feb).
\newblock Neutron drip line in the {Ca} region from {Bayesian Model Averaging}.
\newblock {\em Physical Review Letters\/}~{\em 122}, 062502.

\bibitem[\protect\citeauthoryear{Neufcourt, Cao, Nazarewicz, and
  Viens}{Neufcourt et~al.}{2018}]{Neufcourt2018}
Neufcourt, L., Y.~Cao, W.~Nazarewicz, and F.~Viens (2018).
\newblock Bayesian approach to model-based extrapolation of nuclear
  observables.
\newblock {\em Physical Review C\/}~{\em 98}, 034318.

\bibitem[\protect\citeauthoryear{Pajor}{Pajor}{2017}]{Pajor2017}
Pajor, A. (2017).
\newblock {Estimating the Marginal Likelihood Using the Arithmetic Mean
  Identity}.
\newblock {\em Bayesian Analysis\/}~{\em 12\/}(1), 261 -- 287.

\bibitem[\protect\citeauthoryear{Papamakarios, Nalisnick, Rezende, Mohamed, and
  Lakshminarayanan}{Papamakarios et~al.}{2021}]{JMLR:v22:19-1028}
Papamakarios, G., E.~Nalisnick, D.~J. Rezende, S.~Mohamed, and
  B.~Lakshminarayanan (2021).
\newblock Normalizing flows for probabilistic modeling and inference.
\newblock {\em Journal of Machine Learning Research\/}~{\em 22\/}(57), 1--64.

\bibitem[\protect\citeauthoryear{Papamakarios, Pavlakou, and
  Murray}{Papamakarios et~al.}{2017}]{NIPS2017_6c1da886}
Papamakarios, G., T.~Pavlakou, and I.~Murray (2017).
\newblock Masked autoregressive flow for density estimation.
\newblock In I.~Guyon, U.~V. Luxburg, S.~Bengio, H.~Wallach, R.~Fergus,
  S.~Vishwanathan, and R.~Garnett (Eds.), {\em Advances in Neural Information
  Processing Systems}, Volume~30. Curran Associates, Inc.

\bibitem[\protect\citeauthoryear{Paszke, Gross, Massa, Lerer, Bradbury, Chanan,
  Killeen, Lin, Gimelshein, Antiga, Desmaison, Kopf, Yang, DeVito, Raison,
  Tejani, Chilamkurthy, Steiner, Fang, Bai, and Chintala}{Paszke
  et~al.}{2019}]{Torch}
Paszke, A., S.~Gross, F.~Massa, A.~Lerer, J.~Bradbury, G.~Chanan, T.~Killeen,
  Z.~Lin, N.~Gimelshein, L.~Antiga, A.~Desmaison, A.~Kopf, E.~Yang, Z.~DeVito,
  M.~Raison, A.~Tejani, S.~Chilamkurthy, B.~Steiner, L.~Fang, J.~Bai, and
  S.~Chintala (2019).
\newblock Pytorch: An imperative style, high-performance deep learning library.
\newblock In {\em Advances in Neural Information Processing Systems 32}, pp.\
  8024--8035. Curran Associates, Inc.

\bibitem[\protect\citeauthoryear{Peterson and Anderson}{Peterson and
  Anderson}{1987}]{Peterson}
Peterson, C. and J.~R. Anderson (1987).
\newblock A mean field theory learning algorithm for neural networks.
\newblock {\em Complex Systems\/}~{\em 1}, 995--1019.

\bibitem[\protect\citeauthoryear{Phillips, Furnstahl, Heinz, Maiti, Nazarewicz,
  Nunes, Plumlee, Pratola, Pratt, Viens, and Wild}{Phillips
  et~al.}{2021}]{BAND}
Phillips, D.~R., R.~J. Furnstahl, U.~Heinz, T.~Maiti, W.~Nazarewicz, F.~M.
  Nunes, M.~Plumlee, M.~T. Pratola, S.~Pratt, F.~G. Viens, and S.~M. Wild
  (2021, 5).
\newblock Get on the band wagon: a bayesian framework for quantifying model
  uncertainties in nuclear dynamics.
\newblock {\em Journal of Physics. G, Nuclear and Particle Physics\/}~{\em
  48\/}(7).

\bibitem[\protect\citeauthoryear{Raftery, Madigan, and Hoeting}{Raftery
  et~al.}{1997}]{Raftery1997}
Raftery, A.~E., D.~Madigan, and J.~A. Hoeting (1997).
\newblock Bayesian model averaging for linear regression models.
\newblock {\em Journal of the American Statistical Association\/}~{\em
  92\/}(437), 179--191.

\bibitem[\protect\citeauthoryear{Raftery, Newton, Satagopa, and
  Krivitsk}{Raftery et~al.}{2007}]{Raftery2007}
Raftery, A.~E., M.~A. Newton, J.~M. Satagopa, and P.~N. Krivitsk (2007).
\newblock Estimating the integrated likelihood via posterior simulation using
  the harmonic mean identity.
\newblock In {\em Bayesian Statistics 8}, pp.\  1--45. Oxford University Press.

\bibitem[\protect\citeauthoryear{Ranganath, Gerrish, and Blei}{Ranganath
  et~al.}{2014}]{Ranganath14}
Ranganath, R., S.~Gerrish, and D.~Blei (2014).
\newblock {Black box variational inference}.
\newblock In {\em Proceedings of the Seventeenth International Conference on
  Artificial Intelligence and Statistics}, Volume~33 of {\em Proceedings of
  Machine Learning Research}, pp.\  814--822. PMLR.

\bibitem[\protect\citeauthoryear{Ranganath, Tran, and Blei}{Ranganath
  et~al.}{2016}]{Ranganath:2016}
Ranganath, R., D.~Tran, and D.~M. Blei (2016).
\newblock Hierarchical variational models.
\newblock In {\em Proceedings of the 33rd International Conference on
  International Conference on Machine Learning -- Volume 48}, ICML'16, pp.\
  2568--2577. JMLR.

\bibitem[\protect\citeauthoryear{Rezende and Mohamed}{Rezende and
  Mohamed}{2015}]{pmlr-v37-rezende15}
Rezende, D. and S.~Mohamed (2015, 07--09 Jul).
\newblock Variational inference with normalizing flows.
\newblock In F.~Bach and D.~Blei (Eds.), {\em Proceedings of the 32nd
  International Conference on Machine Learning}, Volume~37 of {\em Proceedings
  of Machine Learning Research}, Lille, France, pp.\  1530--1538. PMLR.

\bibitem[\protect\citeauthoryear{Robbins and Monro}{Robbins and
  Monro}{1951}]{robbins1951}
Robbins, H. and S.~Monro (1951, 09).
\newblock A stochastic approximation method.
\newblock {\em Annals of Mathematical Statistics\/}~{\em 22\/}(3), 400--407.

\bibitem[\protect\citeauthoryear{Ross}{Ross}{2006}]{Ross:2006}
Ross, S.~M. (2006).
\newblock {\em Simulation\/} (Fourth ed.).
\newblock Orlando, FL: Academic Press, Inc.

\bibitem[\protect\citeauthoryear{Ruiz, Titsias, and Blei}{Ruiz
  et~al.}{2016}]{OBBVI}
Ruiz, F. J.~R., M.~K. Titsias, and D.~M. Blei (2016).
\newblock Overdispersed black-box variational inference.
\newblock In {\em Proceedings of the Thirty-Second Conference on Uncertainty in
  Artificial Intelligence}, UAI’16, Arlington, Virginia, USA, pp.\
  647–656. AUAI Press.

\bibitem[\protect\citeauthoryear{Salvatier, Wiecki, and Fonnesbeck}{Salvatier
  et~al.}{2016}]{salvatier2016probabilistic}
Salvatier, J., T.~V. Wiecki, and C.~Fonnesbeck (2016).
\newblock Probabilistic programming in python using pymc3.
\newblock {\em PeerJ Computer Science\/}~{\em 2}, e55.

\bibitem[\protect\citeauthoryear{Schorning, Bornkamp, Bretz, and
  Dette}{Schorning et~al.}{2016}]{med-SCH16}
Schorning, K., B.~Bornkamp, F.~Bretz, and H.~Dette (2016).
\newblock Model selection versus model averaging in dose finding studies.
\newblock {\em Statistics in Medicine\/}~{\em 35}, 4021--4040.

\bibitem[\protect\citeauthoryear{Shazeer and Stern}{Shazeer and
  Stern}{2018}]{pmlr-v80-shazeer18a}
Shazeer, N. and M.~Stern (2018, 10--15 Jul).
\newblock Adafactor: Adaptive learning rates with sublinear memory cost.
\newblock In J.~Dy and A.~Krause (Eds.), {\em Proceedings of the 35th
  International Conference on Machine Learning}, Volume~80 of {\em Proceedings
  of Machine Learning Research}, pp.\  4596--4604. PMLR.

\bibitem[\protect\citeauthoryear{Silvestro, Schnitzler, Liow, Antonelli, and
  Salamin}{Silvestro et~al.}{2014}]{ecol-Si14}
Silvestro, D., J.~Schnitzler, L.~H. Liow, A.~Antonelli, and N.~Salamin (2014).
\newblock Bayesian estimation of speciation and extinction from incomplete
  fossil occurrence data.
\newblock {\em Systematic Biology\/}~{\em 63}, 349--367.

\bibitem[\protect\citeauthoryear{Skilling}{Skilling}{2006}]{NestedSampling}
Skilling, J. (2006).
\newblock Nested sampling for general bayesian computation.
\newblock {\em Bayesian Analysis\/}~{\em 1\/}(4), 833--860.

\bibitem[\protect\citeauthoryear{Steel}{Steel}{2020}]{Steel2020}
Steel, M. F.~J. (2020, September).
\newblock Model averaging and its use in economics.
\newblock {\em Journal of Economic Literature\/}~{\em 58\/}(3), 644--719.

\bibitem[\protect\citeauthoryear{Tieleman and Hinton}{Tieleman and
  Hinton}{2012}]{Tieleman2012}
Tieleman, T. and G.~Hinton (2012).
\newblock {Lecture 6.5---RmsProp: Divide the gradient by a running average of
  its recent magnitude}.
\newblock COURSERA: Neural Networks for Machine Learning.

\bibitem[\protect\citeauthoryear{Titsias and Lázaro-Gredilla}{Titsias and
  Lázaro-Gredilla}{2014}]{Titsias14}
Titsias, M. and M.~Lázaro-Gredilla (2014, 22--24 Jun).
\newblock Doubly stochastic variational bayes for non-conjugate inference.
\newblock In E.~P. Xing and T.~Jebara (Eds.), {\em Proceedings of the 31st
  International Conference on Machine Learning}, Volume~32 of {\em Proceedings
  of Machine Learning Research}, Bejing, China, pp.\  1971--1979. PMLR.

\bibitem[\protect\citeauthoryear{Tran, Blei, and Airoldi}{Tran
  et~al.}{2015}]{Tran2015}
Tran, D., D.~M. Blei, and E.~M. Airoldi (2015).
\newblock Copula variational inference.
\newblock In {\em Proceedings of the 28th International Conference on Neural
  Information Processing Systems - Volume 2}, NeurIPS'15, Cambridge, MA, pp.\
  3564--3572. MIT Press.

\bibitem[\protect\citeauthoryear{Tran, Ranganath, and Blei}{Tran
  et~al.}{2017}]{Tran:2017}
Tran, D., R.~Ranganath, and D.~M. Blei (2017).
\newblock Hierarchical implicit models and likelihood-free variational
  inference.
\newblock In {\em Proceedings of the 31st International Conference on Neural
  Information Processing Systems}, NeurIPS'17, pp.\  5529--5539.

\bibitem[\protect\citeauthoryear{Vandaele}{Vandaele}{1978}]{vandaele1978participation}
Vandaele, W. (1978).
\newblock Participation in illegitimate activities-ehrlich revisited (from
  deterrence and incapacitation-estimating the effects of criminal sanctions on
  crime rates, p 270-335, 1978, alfred blumstein et al, ed.-see ncj-44669).

\bibitem[\protect\citeauthoryear{Vaswani, Shazeer, Parmar, Uszkoreit, Jones,
  Gomez, Kaiser, and Polosukhin}{Vaswani et~al.}{2017}]{NIPS2017_3f5ee243}
Vaswani, A., N.~Shazeer, N.~Parmar, J.~Uszkoreit, L.~Jones, A.~N. Gomez, L.~u.
  Kaiser, and I.~Polosukhin (2017).
\newblock Attention is all you need.
\newblock In I.~Guyon, U.~V. Luxburg, S.~Bengio, H.~Wallach, R.~Fergus,
  S.~Vishwanathan, and R.~Garnett (Eds.), {\em Advances in Neural Information
  Processing Systems}, Volume~30. Curran Associates, Inc.

\bibitem[\protect\citeauthoryear{Wainwright and Jordan}{Wainwright and
  Jordan}{2008}]{MAL001}
Wainwright, M.~J. and M.~I. Jordan (2008).
\newblock Graphical models, exponential families, and variational inference.
\newblock {\em Foundations and Trends® in Machine Learning\/}~{\em
  1\/}(1–2), 1--305.

\bibitem[\protect\citeauthoryear{Wang, Audi, Kondev, Huang, Naimi, and Xu}{Wang
  et~al.}{2017}]{AME2016}
Wang, M., G.~Audi, F.~G. Kondev, W.~J. Huang, S.~Naimi, and X.~Xu (2017).
\newblock The {{AME2016}} atomic mass evaluation ({II}). tables, graphs and
  references.
\newblock {\em Chinese Physics C\/}~{\em 41}, 030003.

\bibitem[\protect\citeauthoryear{Wang and Blei}{Wang and Blei}{2018}]{VIth}
Wang, Y. and D.~M. Blei (2018).
\newblock Frequentist consistency of variational {B}ayes.
\newblock {\em Journal of the American Statistical Association\/}~{\em 0\/}(0),
  1--15.

\bibitem[\protect\citeauthoryear{Wei, Visweswaran, and Cooper}{Wei
  et~al.}{2011}]{gen-Vis11}
Wei, W., S.~Visweswaran, and G.~F. Cooper (2011).
\newblock The application of naive {{Bayes}} model averaging to predict
  {{Alzheimer}}'s disease from genome-wide data.
\newblock {\em Journal of the American Medical Informatics Association\/}~{\em
  18}, 370--375.

\bibitem[\protect\citeauthoryear{Weilbach, Beronov, Wood, and Harvey}{Weilbach
  et~al.}{2020}]{pmlr-v108-weilbach20a}
Weilbach, C., B.~Beronov, F.~Wood, and W.~Harvey (2020, 26--28 Aug).
\newblock Structured conditional continuous normalizing flows for efficient
  amortized inference in graphical models.
\newblock In S.~Chiappa and R.~Calandra (Eds.), {\em Proceedings of the Twenty
  Third International Conference on Artificial Intelligence and Statistics},
  Volume 108 of {\em Proceedings of Machine Learning Research}, pp.\
  4441--4451. PMLR.

\bibitem[\protect\citeauthoryear{Wen}{Wen}{2015}]{gen-Wen15}
Wen, X. (2015).
\newblock Bayesian model comparison in genetic association analysis: linear
  mixed modeling and {{SNP}} set testing.
\newblock {\em Biostatistics\/}~{\em 16}, 701--712.

\bibitem[\protect\citeauthoryear{Zeiler}{Zeiler}{2012}]{Zeiler2012}
Zeiler, M.~D. (2012).
\newblock Adadelta: An adaptive learning rate method.
\newblock {\em ArXiv\/}~{\em 1212.5701}.

\bibitem[\protect\citeauthoryear{Zellner}{Zellner}{1986}]{Zellner1986}
Zellner, A. (1986).
\newblock On assessing prior distributions and bayesian regression analysis
  with g prior distributions.
\newblock {\em Bayesian Inference and Decision Techniques: Essays in Honor of
  Brune de Finetti\/}, 233--243.

\bibitem[\protect\citeauthoryear{Zhang and Gao}{Zhang and Gao}{2020}]{ZhangGao}
Zhang, F. and C.~Gao (2020).
\newblock {Convergence rates of variational posterior distributions}.
\newblock {\em The Annals of Statistics\/}~{\em 48\/}(4), 2180 -- 2207.

\end{thebibliography}
\end{document}